\documentclass[10pt,twocolumn]{article}

\usepackage{usenix2019}
\usepackage{xcolor}
\pdfoutput=1

\usepackage{times}
\usepackage{fullpage}

\newcommand{\name}{numpywren}
\newcommand{\irname}{LAmbdaPACK}
\usepackage{graphicx}
\usepackage{algorithm}
\usepackage{algorithmic}
\usepackage{etoolbox}
\usepackage{subcaption}
\usepackage{authblk}

\usepackage{microtype}

\usepackage{hyperref}
\usepackage{url}
\usepackage[frozencache]{minted}
\usepackage{booktabs}
\usepackage{amsmath}

\newcommand{\algorithmicdoinparallel}{\textbf{do in parallel}}
\makeatletter
\AtBeginEnvironment{algorithmic}{%
  \newcommand{\FORALLP}[2][default]{\ALC@it\algorithmicforall\ #2\ %
    \algorithmicdoinparallel\ALC@com{#1}\begin{ALC@for}}%
}
\makeatother

\newlength\myindent
\graphicspath{ {images/} }

\begin{document}

\title{\bf {\name}: Serverless Linear Algebra }

\author[1]{ Vaishaal Shankar}
\author[1]{Karl Krauth}
\author[1]{Qifan Pu}
\author[1]{\\Eric Jonas}
\author[2]{Shivaram Venkataraman}
\author[1]{ Ion Stoica}
\author[1]{Benjamin Recht}
\author[1]{Jonathan Ragan-Kelley}

\affil[1]{\footnotesize UC Berkeley}
\affil[2]{\footnotesize UW Madison}

\date{}
\maketitle
\thispagestyle{empty}

\begin{abstract}

Linear algebra operations are widely used in scientific computing and machine learning applications. However, it is challenging for scientists and data analysts to run linear algebra at scales beyond a single machine. Traditional approaches either require access to supercomputing clusters, or impose configuration and cluster management challenges. In this paper we show how the disaggregation of storage and compute resources in so-called ``serverless'' environments, combined with compute-intensive workload characteristics, can be exploited to achieve elastic scalability and ease of management. 

We present {\name}, a system for linear algebra built on a serverless architecture. We also introduce {\irname}, a domain-specific language designed to implement highly parallel linear algebra algorithms in a serverless setting. We show that, for certain linear algebra algorithms such as matrix multiply, singular value decomposition, and Cholesky decomposition, {\name}'s performance (completion time) is within 33\% of ScaLAPACK, and its compute efficiency (total CPU-hours) is up to 240\% better due to elasticity, while providing an easier to use interface and better fault tolerance.
At the same time, we show that the inability of serverless runtimes to exploit locality \emph{across} the cores in a machine fundamentally limits their network efficiency, which limits performance on other algorithms such as QR factorization.
This highlights how cloud providers could better support these types of computations through small changes in their infrastructure.  %
\end{abstract}

\vspace{-0.8em}
\section{Introduction}
\vspace{-0.7em}

\begin{figure}[th]
    \centering
    \includegraphics[width=1\linewidth]{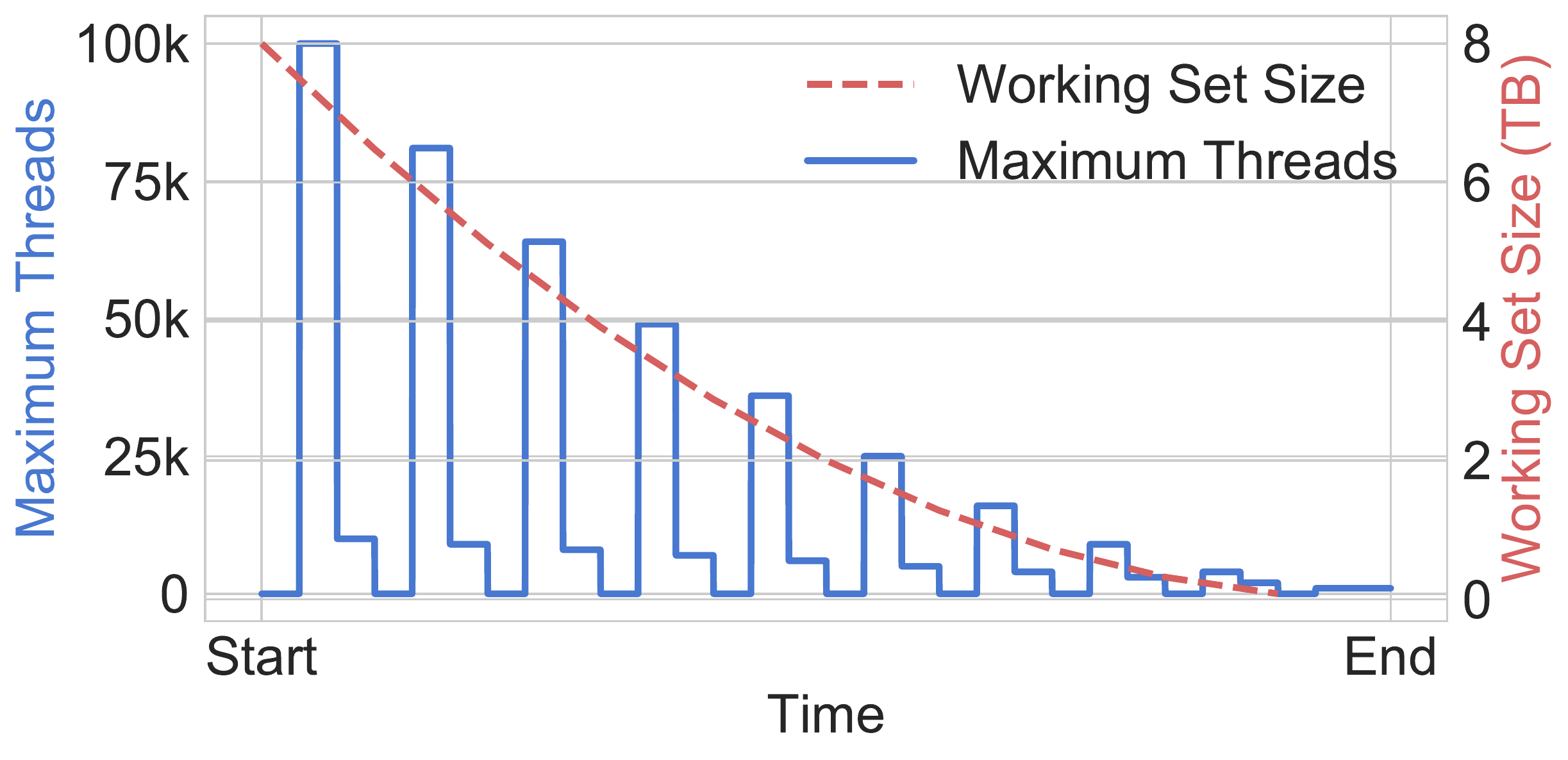}
    \vspace{-2em}
    \caption{ Theoretical profile of available parallelism and required working set size over time in a distributed Cholesky decomposition. Traditional HPC programming models like MPI couple machine parallelism and memory capacity, and require a static allocation for the lifetime of a process.
    This is inefficient both due to the changing ratio of parallelism to working set, and the sharp decrease in utilization over time.}

    \label{fig:theoretical_flops}
\end{figure}

As cloud providers push for resource consolidation and disaggregation~\cite{gao2016network}, we see a shift in distributed computing towards greater elasticity. One such example is the advent of \emph{serverless computing} (e.g.,  AWS Lambda, Google Cloud Functions, Azure Functions) which provides users with instant access to large compute capability without the overhead of managing a complex cluster deployment. While serverless platforms were originally intended for  event-driven, stateless functions, and come with corresponding constraints (e.g., small memory and short run-time limits per invocation), recent work has exploited them for other applications like parallel data analysis~\cite{jonas2017occupy} and distributed video encoding~\cite{fouladi2017encoding}. These workloads are a natural fit for serverless computing as they are either embarrassingly parallel or use simple communication patterns across functions. Exactly how complex the communication patterns and workloads can be and still efficiently fit in a stateless framework remains an active research question.\\ 
Linear algebra operations are at the core of many data-intensive applications.
Their wide applicability covers both traditional scientific computing problems such as weather simulation, genome assembly, and fluid dynamics, as well as emerging computing workloads, including distributed optimization~\cite{parikh2014proximal}, robust control~\cite{tu2017non} and computational imaging~\cite{heide2016proximal}. As the data sizes for these problems continue to grow, we see increasing demand for running linear algebra computations at large scale. 

Unfortunately, running large-scale distributed linear algebra remains challenging for many scientists and data analysts due to accessibility, provisioning, and cluster management constraints. Traditionally, such linear algebra workloads are run on managed high performance computing (HPC) clusters, access to which is often behind walls of paperwork and long job wait queues. To lower the bar for access, providers such as Amazon Web Services (AWS), Google, and Microsoft Azure now provide HPC clusters in the cloud~\cite{awshpc, googlehpc, azurehpc}.
While the HPC-in-the-cloud model looks promising, it adds extra configuration complexity, since users have to choose from a complex array of configuration options including cluster size, machine types, and storage types~\cite{ernest}.

This extends to many existing systems that run large-scale linear algebra on data parallel systems~\cite{madlinq,marlin,systemml} and that are deployed on a cluster of virtual machines (VMs).
This complexity is further exacerbated by the fact that many linear algebra workloads have large dynamic range in memory and computation requirements over the course of their execution. For example, performing Cholesky decomposition~\cite{ballard2010communication}---one of the most popular methods for solving systems of linear equations---on a large matrix generates computation phases with oscillating parallelism and decreasing working set size (Figure~\ref{fig:theoretical_flops}). Provisioning a cluster of any static size will either slow down the job or leave the cluster under-utilized.

Our key insight is that, for many linear algebra operations, regardless of their complex structure, computation time often dominates communication for large problem sizes, e.g., $O(n^3)$ compute and $O(n^{2})$ communication for Cholesky decomposition. Thus, with appropriate blocking and pipelining, we find that it is possible to use high-bandwidth but high-latency distributed \emph{storage} as a substitute for large-scale distributed \emph{memory}.

Based on this idea, we design {\name}, a system for linear algebra on serverless architectures. {\name} runs computations as stateless functions while storing intermediate state in a distributed object store.
{\name} executes programs written using {\irname}, a high level DSL we designed that makes it easy to express state-of-the-art communication avoiding linear algebra algorithms~\cite{anderson2011communication} with fine-grained parallelism. Importantly, operating on large matrices at fine granularity can lead to very large task graphs (16M nodes for a matrix with 1M rows and columns, even with a relatively coarse block size of 4K), and the lack of a dedicated driver in the serverles setting would mean each worker would need a copy of the task graph to reason about the dependencies in the program. We address this by using ideas from the literature of loop optimization and show that the {\irname} runtime can scale to large matrix sizes while generating programs of constant size.

Our evaluation shows that for a number of important linear algebra operations (e.g., Cholesky decomposition, matrix multiply, SVD) {\name} can rival the performance of highly optimized distributed linear algebra libraries running on a dedicated cluster.  
We also show that in these favorable cases {\name} is more flexible and can consume $32\%$ fewer CPU-hours, while being fault-tolerant. Compared to fault-tolerant data parallel systems like Dask, we find that {\name} is up to 320\% faster and can scale to larger problem sizes. We also show that with \irname{} we can implicitly represent  structured task graphs with millions of nodes in as little as 2 KB.

However, for \textbf{all} algorithms stateless function execution imposes large communication overhead. Most distributed linear algebra algorithms heavily exploit locality where an instance with $n$ cores can share a single copy of the data. In serverless systems, every a function has a single core and as these functions could be execute on any machine, we need to send $n$ copies of the data to reach $n$ cores. These limitations affect our performance for certain algorithms, such as QR decomposition. We discuss these limitations and potential solutions in Sec~\ref{sec:exps}.

In summary we make the following contributions:
\begin{enumerate}
\item We provide the first concrete evidence that certain large scale linear algebra algorithms can be efficiently executed using purely stateless functions and disaggregated storage.
\item We design {\irname}, a domain specific language for linear algebra algorithms that captures fine grained dependencies and can express state of the art communication avoiding linear algebra algorithms in a succinct and readable manner.
\item We show that {\name} can scale to run Cholesky decomposition on a 1Mx1M matrix, and is within 36\% of the completion time of ScaLAPACK running on dedicated instances, and can be tuned to use 33\% fewer CPU-hours.

\end{enumerate}

\vspace{-1em}
\section{Background}
\vspace{-0.8em}
\label{sec:background}
\label{sec:serverless}

\subsection{Serverless Landscape}
\vspace{-0.3em}

In the serverless computing model, cloud providers offer the ability to execute functions on demand, hiding cluster configuration and management overheads from end users.
In addition to the usability benefits, this model also improves efficiency: the cloud provider can multiplex resources at a much finer granularity than what is possible with traditional cluster computing, and the user is not charged for idle resources. However, in order to efficiently manage resources, cloud providers place limits on the use of each resource. We next discuss how these constraints affect the design of our system. 

\noindent\textbf{Computation.}
Computation resources offered in serverless platforms are typically restricted to a single CPU core and a short window of computation.
For example AWS Lambda provides 300 seconds of compute on a single AVX/AVX2 core with access to up to 3 GB of memory and 512 MB of disk storage. 
Users can execute a number of parallel functions, and, as one would expect, the aggregate compute performance of these executions scales almost linearly.

The linear scalability in function execution is only useful for embarrassingly parallel computations when there is no communication between the individual workers.
Unfortunately, as individual workers are transient and as their start-up times could be staggered, a traditional MPI-like model of peer-to-peer communication will not work in this environment.
This encourages us to leverage storage, which can be used as an indirect communication channel between workers.

\noindent\textbf{Storage.}
Cloud providers offer a number of storage options ranging from key-value stores to relational databases. Some services are not purely 
elastic in the sense that they require resources to be provisioned beforehand. However distributed object storage systems like Amazon S3 or Google Cloud Storage
offer unbounded storage where users are only charged for the amount of data stored.
From the study done in~\cite{jonas2017occupy} we see that AWS Lambda function invocations can read and write to Amazon S3 at 250 GB/s. Having access to such high bandwidth means that
we can potentially store intermediate state during computation in a distributed object store. However such object stores typically have high latency ($\sim$10ms) to access any key meaning
we need to design our system to perform coarse-grained access.
Finally, the cost of data storage in an object storage system is often orders of magnitude lower when compared to instance memory. For example on Amazon S3 the price of data storage is \$0.03 per TB-hour; in contrast the cheapest large memory instances are priced at \$6 per TB-hour. This means that using a storage system could be cheaper if the access pattern does not require instance memory. 

\noindent\textbf{PubSub.}
In addition to storage services, cloud providers also offer publish-subscribe services like Amazon SQS or Google Task Queue. These services typically do not support high data bandwidths
but can be used for ``control plane" state like a task queue that is shared between all serverless function invocations. Providers often offer consistency guarantees for these services, and most services guarantee at least once delivery.

\vspace{-0.6em}

\vspace{-0.6em}
\subsection{Linear Algebra Algorithms}
\vspace{-0.4em}

Given the motivating applications, in this work, we broadly focus on the case of large-scale \emph{dense} linear algebra.
Algorithms in this regime have a rich literature of parallel communication-avoiding algorithms and existing high performance implementations~\cite{anderson2011communication, ballard2010communication, ballard2011minimizing, georganas2012communication}. 
To motivate the design decisions in the subsequent sections we briefly review the communication and computation patterns of a core subroutine in solving a linear system, Cholesky factorization. \\
\noindent\textbf{Case study:}
Cholesky factorization is one of the most popular algorithms for solving linear equations, and it is widely used in applications such as matrix inversion, partial differential equations, and Monte Carlo simulations.
To illustrate the use of Cholesky decomposition, consider the problem of solving a linear equation $Ax = b$, where A is a symmetric positive definite matrix. One can first perform a Cholesky decomposition of $A$ into two triangular matrices $A = LL^{T}$ ($\mathcal{O}(n^3)$), then solve two relatively simpler equations of $Ly=b$ ($\mathcal{O}(n^2)$ via forward substitution) and $L^{T}x=y$ ($\mathcal{O}(n^2)$ via back substitution) to obtain the solution $x$. From this process, we can see that the decomposition is the most expensive step.

\emph{Communication-Avoiding Cholesky}~\cite{ballard2010communication} is a well-studied routine to compute a Cholesky decomposition. The algorithm divides the matrix into blocks and derives a computation order that minimizes total data transfer. We pick this routine not only because it is one of the most performant, but also because it showcases the structure of computation found in many linear algebra algorithms. %

\begin{algorithm}[!t]%
   \textbf{Input:} \\ 
  $A$ - Positive Semidefinite Symmetric Matrix \\
  $B$ - block size \\
  $N$ - number of rows in $A $\\
   \textbf{Blocking:} \\
   $A_{ij}$ - the $ij$-th block of A \\
   \textbf{Output:} \\
   $L$ - Cholesky Decomposition of $A$ \\
\begin{algorithmic}[1] %
   \FOR{$j \in \left\{ 0...\lceil \frac{N}{B} \rceil \right\}$}
    \STATE $L_{jj} \Leftarrow cholesky(A_{jj})$
    \FORALLP{$i \in \left\{ j+1...\lceil \frac{N}{B} \rceil \right\}$}
     \STATE {$L_{ij} \Leftarrow L_{jj}^{-1} A_{ij}$}
    \ENDFOR \\
     \FORALLP{$k \in \left\{ j+1...\lceil \frac{N}{B} \rceil \right\}$}
        \FORALLP{$l \in \left\{ k...\lceil \frac{N}{B} \rceil \right\}$}
          \STATE {$A_{kl} \Leftarrow A_{kl} - L_{kj}^{T}L_{lj}$}
             \ENDFOR
         \ENDFOR
   \ENDFOR 
\end{algorithmic}
\caption{Communication-Avoiding Cholesky \cite{ballard2010communication}} %
\label{alg:chol}
\end{algorithm}

\begin{figure*}[!t]
    \includegraphics[width=0.92\textwidth]{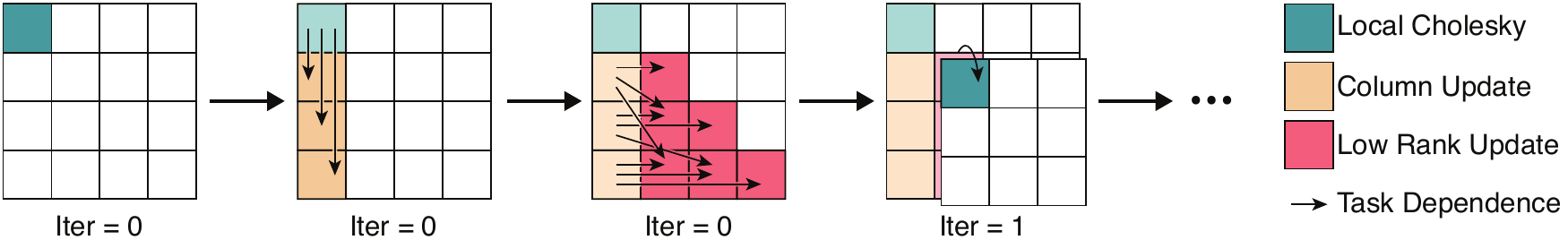}
    \caption{First 4 time steps of parallel Cholesky decomposition: 0) Diagonal block Cholesky decomposition 1) Parallel column update 2) Parallel submatrix update 3) (subsequent) Diagonal block Cholesky decomposition}
    \label{fig:cholesky}
    \vspace{-0.1in}
\end{figure*}

The pseudo-code for communication-avoiding Cholesky decomposition is shown in Algorithm~\ref{alg:chol}. At each step of  the outer loop ($j$), the algorithm first computes Cholesky decomposition of a single block $A_{jj}$ (Fig.~\ref{fig:cholesky}(a)). This result is used to update the ``panel'' consisting of the column blocks below $A_{ij}$ (Fig.~\ref{fig:cholesky}(b)). Finally all blocks to the right of column $j$ are updated by indexing the panel according to their respective positions (Fig.~\ref{fig:cholesky}(c)). This process is repeated by moving down the diagonal (Fig.~\ref{fig:cholesky}(d)).

We make two key observations from analyzing the computational structure of Algorithm~\ref{alg:chol}. First, we see that the algorithm exhibits {\em dynamic parallelism} during execution. The outer loop consists of three distinct steps with different amounts of parallelism, from $\mathcal{O}(1)$, $\mathcal{O}(K)$ to $\mathcal{O}(K^2)$, where $K$ is the enclosing sub-matrix size at each step. In addition, as $K$ decreases at each iteration, overall parallelism available for each iteration decreases from $\mathcal{O}(K^2)$ to $\mathcal{O}(1)$ as shown in Figure~\ref{fig:theoretical_flops}. 
Our second observation is that the algorithm has {\em fine-grained dependencies} between the three steps, both within an iteration and across iterations. For example, $A_{kl}$ in step 3 can be computed as long as $L_{kj}$ and $L_{lj}$ are available (line 8). Similarly, the next iteration can start as soon as $A_{(j+1)(j+1)}$ is updated. Such fine-grained dependencies are hard to exploit in single program multiple data (SPMD) or bulk synchronous parallel (BSP) systems  such as MapReduce or Apache Spark, where global synchronous barriers are enforced between steps.

\subsection{{\name} Overview}

We design {\name} to target linear algebra workloads that have execution patterns similar to Cholesky decomposition described above. Our goal is to adapt to the amount of parallelism when available and we approach this by decomposing programs into fine-grained execution units that can be run in parallel. To achieve this at scale in a stateless setting, we propose performing dependency analysis in a \emph{decentralized} fashion. We distribute a global dependency graph describing the control flow of the program to every worker. Each worker then locally reasons about its down stream dependencies based on its current position in the global task graph. In the next two sections we will describe {\irname} the DSL that allows for compact representations of these global dependency graphs, and the {\name} execution engine that runs the distributed program. 

\vspace{-0.2in}
\section{Programming Model}
\label{sec:comp}
\vspace{-0.1in}
In this section we present an overview of \irname{}, our domain specific language for specifying parallel linear algebra algorithms.
Classical algorithms for high performance linear algebra are difficult to map directly to a serverless environment
as they rely heavily on peer-to-peer communication and exploit locality of data and computation -- luxuries absent in a serverless computing cluster.
Furthermore, most existing implementations of linear algebra algorithms like ScalaPACK are explicitly designed for stateful HPC clusters.

We thus design \irname{} to adapt ideas from recent advances in the numerical linear algebra community on expressing algorithms as directed acyclic graph (DAG) based computation~\cite{plasma, dplasma}. Particularly \irname{} borrows techniques from Dague \cite{dague} a  DAG execution framework aimed at HPC environments, though we differ in our analysis methods and target computational platform.
We design \irname{} to allow users to succinctly express \textit{tiled} linear algebra algorithms. These routines express their computations as operations on matrix \textit{tiles}, small submatrices that can fit in local memory. The main distinction between tiled algorithms and the classical algorithms found in libraries like ScaLAPACK is that the algorithm itself is agnostic to machine layout, connectivity, etc., and only defines a computational graph on the block indices of the matrices.
This uniform, machine independent abstraction for defining complex algorithms allows us to adapt most standard linear algebra routines to a stateless execution engine.

\vspace{-0.1in}
\subsection{Language Design}

\irname{} programs are simple imperative routines which produce and consume tiled matrices.
These programs can perform basic arithmetic and logical operations on scalar values.
They cannot directly read or write matrix values;
instead, all substantive computation is performed by calling native kernels on matrix tiles.
Matrix tiles are referenced by index, and the primary role of the \irname{} program is to
sequence kernel calls, and to compute the tile indices for each.\\
\irname{} programs include simple for loops and if statements, but there is no recursion, only a single level of function calls, from the \irname{} routine to kernels.
Each matrix tile index can be written to only once, a common design principle in many functional languages \footnote{Arbitrary programs can be easily translated into this static single assignment form, but we have found it natural to program directly in this style}.
Capturing index expressions as symbolic objects in this program is key to the dependence analysis we perform.\\
These simple primitives are powerful enough to concisely implement algorithms such as Tall Skinny QR (TSQR), LU, Cholesky, and Singular Value decompositions.
A  description of \irname{} is shown in Figure~\ref{fig:types},
and examples of concrete \irname{} implementations of  Cholesky and TSQR are shown in Figures ~\ref{fig:cholesky_alg} and ~\ref{fig:tsqr}.

\begin{figure}[ht]
    \begin{minted}[fontsize=\footnotesize,xleftmargin=1pt]{ocaml}
Uop = Neg| Not| Log| Ceiling| Floor| Log2
Bop = Add| Sub| Mul| Div| Mod| And| Or
Cop = EQ | NE | LT | GT | LE | GE
    
IdxExpr = IndexExpr(Str matrix_name,
            Expr[] indices)
     
Expr = BinOp(Bop op, Expr left, Expr right)
       | CmpOp(Cop op, Expr left, Expr right)
       | UnOp(Uop op, Expr e)
       | Ref(Str name)
       | FloatConst(float val)
       | IntConst(int val)

Stmt = KernelCall(Str fn_name,
         IdxExpr[] outputs,
         IdxExpr[] matrix_inputs,
         Expr[] scalar_inputs)
       | Assign(Ref ref, Expr val)
       | Block(Stmt* body)
       | If(Expr cond, Stmt body, Stmt? else)
       | For(Str var, Expr min, 
           Expr max, Expr step, Stmt body)
    \end{minted}
    \vspace{-0.2in}
    \caption{\label{fig:types} A description of the \irname{} language.}
\end{figure}

\subsection{Program Analysis}
\label{sec:analysis}
There are no parallel primitives present in \irname{}, but rather the \irname{} runtime deduces the underlying dependency graph by statically analyzing the program.
In order to execute a program in parallel, we construct a DAG of kernel calls from the dependency structure induced by the program.
Naively converting the program into an executable graph will lead to a \emph{DAG explosion} as the size of the data structure required to represent the program will scale with the size of the input \emph{data}, which can lead to intractable compilation times.
Most linear algebra algorithms of interest are $\mathcal{O}(N^{3})$, and even fast symbolic enumeration of $\mathcal{O}(N^{3})$ operations at runtime as used by systems like MadLINQ~\cite{madlinq} can lead to intractable compute times and overheads for large problems.

In contrast, we borrow and extend techniques from the loop optimization community to convert a \irname{} program into an \emph{implicit} directed acyclic graph. 
We represent each node $\mathcal{N}$ in the program's DAG as a tuple of \verb|(line_number, loop_indices)|. With this information any statement in the program's iteration space can be executed.
The challenge now lies in deducing the downstream dependencies given a particular node in the DAG.
Our approach is to handle dependency analysis at \emph{at runtime}: whenever a storage location is being written to, we determine expressions in $\mathcal{N}$ (all lines, all loop indices) that read from the same storage location.

We solve the problem of determining downstream dependencies for a particular node by modeling the constraints as a system of equations. 
We assume that the number of lines in a single linear algebra algorithm will be necessarily small.
However, the iteration space of the program can often be far too large to enumerate directly (as mentioned above, this is often as large as $\mathcal{O}(n^{3})$).
Fortunately the pattern of data accesses in linear algebra algorithms is highly structured.
Particularly when arrays are indexed solely by \emph{affine functions of loop variables}---that is functions of the form $ai + b$, where $i$ is a loop variable and $a$ and $b$ are constants known at compile time---standard techniques from loop optimization can be employed to efficiently find the dependencies of a particular node. These techniques often involve solving a small system of integer-valued linear equations, where the number of variables in the system depends on the number of nested loop variables in the program.\\
\noindent\textbf{Example of linear analysis.}
Consider the Cholesky program in Figure~\ref{fig:cholesky_alg}. If at runtime a worker is executing line 7 of the program with $i=0$, $j=1$ and $k=1$, to find the downstream dependencies, the analyzer will scan each of the 7 lines of the program and calculate whether there exists a valid set of loop indices such that $S[1,1,1]$ can be read from at that point in the program. If so then the tuple of (\verb|line_number|, \verb|loop_indices|) defines the downstream dependency of such task, and becomes a child of the current task. All index expressions in this program contain only affine indices, thus each system can be solved exactly. In this case the only child is the node (2, $\left\{i: 1, j:1, k:1\right\}$). Note that this procedure only depends on the size of the \textbf{program} and not the size of the data being processed.\\
\noindent\textbf{Nonlinearites and Reductions.}
Certain common algorithmic patterns---particularly reductions---involve nonlinear loop bounds and array indices.
Unlike traditional compilers, since all our analysis occurs \emph{at runtime}, all loop boundaries have been determined. Thus we can solve the system of linear and nonlinear equations by first solving the linear equations and using that solution to solve the remaining nonlinear equations. \\
\noindent\textbf{Example of nonlinear analysis.}
Consider the TSQR program in Figure ~\ref{fig:tsqr}. Suppose at runtime a worker is executing line 6 with $level=0$ and $i=6$, then we want to solve for the  loop variable assignments for 
$R[i+2^{level}, level] = R[6,1]$ (line 7). In this case one of the expressions contains a nonlinear term involving $i$ and $level$ and thus we cannot solve for both variables directly. However we can solve for $level$ easily and obtain the value 1. 
We then plug in the resulting value into the nonlinear expression to get a linear equation only involving $i$. 
Then we can solve for $i$ and arrive at the solution (6, $\left\{i:4, level: 1\right\}$). We note that the for loop structures defined by Figure~\ref{fig:tsqr} define a tree reduction with branching factor of 2. 
Using this approach we can capture the nonlinear array indices induced by tree reductions in algorithms such as Tall-Skinny QR (TSQR), Communication Avoiding QR (CAQR), Tournament Pivoting LU (TSLU), and Bidiagonal Factorization (BDFAC). 
The full pseudo code for our analysis algorithm can be found in Algorithm~\ref{alg:analysis}. \\
\noindent\textbf{Implementation.}
To allow for accessibility and ease of development we embed our language in Python. Since most \irname{} call into optimized BLAS and LAPACK kernels, the performance penalty of using a high level interpreted language is small.

\begin{figure}[!t]
    \begin{minted}[fontsize=\footnotesize, xleftmargin=5pt, numbersep=1pt,linenos]{python}
  def cholesky(O:BigMatrix,S:BigMatrix,N:int):
    for i in range(0,N)
      O[i,i] = chol(S[i,i,i])
      for j in range(i+1,N):
        O[j,i] = trsm(O[i,i], S[i,j,i])
        for k in range(i+1,j+1):
          S[i+1,j,k] = syrk(
            S[i,j,k], O[j,i], O[k,i])
    \end{minted}
    \vspace{-0.2in}
    \caption{\label{fig:cholesky_alg} Sample {\irname} of Cholesky Decomposition}
    \vspace{-0.05in}
\end{figure}

\begin{figure}[!t]
    \begin{minted}[fontsize=\footnotesize,xleftmargin=5pt, numbersep=1pt, linenos]{python}
  def tsqr(A:BigMatrix, R:BigMatrix, N:Int):
    for i in range(0,N):
      R[i, 0] = qr_factor(A[i])
    for level in range(0,log2(N))
      for i in range(0,N,2**(level+1)):
        R[i, level+1] = qr_factor(
          R[i, level], R[i+2**level, level])
    \end{minted}
    \vspace{-0.2in}
    \caption{\label{fig:tsqr} Sample {\irname} of Tall-Skinny QR Decomposition}
    \vspace{-0.05in}
\end{figure}

\begin{algorithm}[!t]%
   \textbf{Input:} \\ 
   $\mathcal{P}$ - The source of a \irname{} program \\
   $A$ - a concrete array that is written to \\
   $idx$ - the concrete array index of $A$ writen to \\
   \textbf{Output:} \\
   $O = \{\mathcal{N}_{0}, ..., \mathcal{N}_{k}\}$ - A concrete set of program nodes that read from $A[idx]$ \\
\begin{algorithmic}[1] %
    \STATE{$O = \left\{\right\}$}
    \FOR{$line \in \mathcal{P}$}
        \FOR {$M \in line.read\_matrices$} 
        \IF {$M = A$}
        \STATE{$S = SOLVE(M.symbolic\_idx - idx = 0)$}
        \STATE{$O = O \cup S$}
        \ENDIF
        \ENDFOR
   \ENDFOR
\end{algorithmic}
\caption{{\irname} Analysis}
\label{alg:analysis}
\end{algorithm}

\vspace{-0.2in}
\section{System Design}
\vspace{-0.1in}
\label{sec:arch}

\begin{figure*}[!t]
\centering
\includegraphics[width=1.0\textwidth]{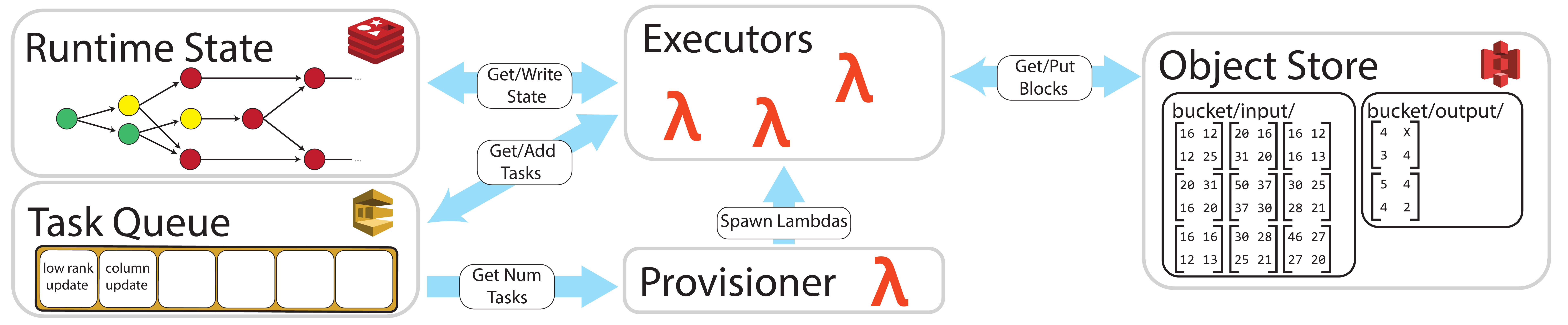}
\caption{\label{fig:execution} The architecture of the execution framework of \name{} showing the runtime state during a $6 \times 6$ cholesky decomposition. The first block cholesky instruction has been executed as well as a single column update.}
\vspace{-0.1in}
\end{figure*}

We next present the system architecture of {\name}. We begin by introducing the high level components in {\name} and trace the execution flow for a computation. Following that we describe techniques to achieve fault tolerance and mitigate stragglers. Finally we discuss the dynamic optimizations that are enabled by our design.

To fully leverage the elasticity and ease-of-management of the cloud, we build {\name} entirely upon existing cloud services while ensuring that we can achieve the performance and fault-tolerance goals for high performance computing workloads. Our system design consists of five major components that are independently scalable: a runtime state store, a task queue, a lightweight global task scheduler, a serverless compute runtime, and a distributed object store. Figure ~\ref{fig:execution} illustrates the components of our system.

The execution proceeds in the following steps:

\noindent\textbf{1. Task Enqueue:} The client process enqueues the first task that needs to be executed into the \emph{task queue}. The task queue is a publish-subscribe style queue that contains all the nodes in the DAG whose input dependencies have been met and are ready to execute. 

\noindent\textbf{2. Executor Provisioning:} The length of the task queue is monitored by a \emph{provisioner} that manages compute resources to match the dynamic parallelism during execution. After the first task is enqueued, the provisioner launches an \emph{executor}. The exact number of stateless workers that are provisioned depends on the \textit{auto-scaling policy} and we discuss the policy used in Section~\ref{subsec:optimizations}. As the provisioner's role is only lightweight it can also be executed periodically as a ``serverless" cloud function.

\noindent\textbf{3. Task Execution:} Executors manage executing and scheduling {\name} tasks. Once an executor is ready, it polls the task queue to fetch the highest priority task available and executes the instructions encoded in the task. Most tasks involve reading input from and writing output to the \emph{object store}, and executing BLAS/LAPACK functions. The object store is assumed to be a distributed, persistent storage system that supports read-after-write consistency for individual keys. Using a persistent object store with a single static assignment language is helpful in designing our fault tolerance protocol. Executors self terminate when they near the runtime limit imposed by many serverless systems (300s for AWS Lambda). The provisioner is then left in charge of launching new workers if necessary. As long as we choose the coarsness of tasks such that many tasks can be successfully completed in the allocated time interval, we do not see too large of a performance penalty for timely worker termination. Our fault tolerance protocol keeps running programs in a valid state even if workers exceed the runtime limit and are killed mid-execution by the cloud provider.

\noindent\textbf{4. Runtime State Update:} Once the task execution is complete and the output has been persisted, the executor updates the task status in the \emph{runtime state store}. The runtime state store tracks the control state of the entire execution and needs to support fast, atomic updates for each task.  If a completed task has children that are ``ready" to be executed the executor adds the child tasks to the task queue. The atomicity of the state store guarantees every child will be scheduled. We would like to emphasize that we only need transactional semantics within the runtime state store, we do not need the runtime state store and the child task enqueuing to occur atomically. We discuss this further in Section~\ref{subsec:ft}. This process of using executors to perform scheduling results in efficient, decentralized, fine grained scheduling of tasks.

\vspace{-0.05in}
\subsection{Fault Tolerance}
\label{subsec:ft}

Fault tolerance in {\name} is much simpler to achieve due to the disaggregation of compute and storage. Because all writes to the object store are made durable, no recomputation is needed after a task is finished. Thus fault tolerance in {\name} is reduced to the problem of recovering failed tasks, in contrast to many systems where all un-checkpointed tasks have to be re-executed~\cite{madlinq}. There are many ways to detect and re-run failed tasks. In {\name} we do this via a simple lease mechanism~\cite{cheriton-leases}, which allows the system to track task status without a scheduler periodically communicating with executors. %

\noindent\textbf{Task Lease:} In {\name}, all the pending and executable tasks are stored in a task queue. We maintain a invariant that a task can only be deleted from the queue once it is completed (i.e., the runtime state store has been updated and the output persisted to the object store). When a task is fetched by a worker, the worker obtains a lease on the task. For the duration of the lease, the task is marked invisible to prevent other workers from fetching the same task. As the lease length is often set to a value that is smaller than task execution time, e.g., 10 seconds, a worker also is responsible for renewing the lease and keeping a task invisible when executing the task.

\noindent\textbf{Failure Detection and Recovery:} During normal operation, the worker will renew lease of the task using a background thread until the task is completed. If the task completes, the worker deletes the task from the queue. If the worker fails, it can no longer renew the lease and the task will become visible to any available workers. Thus, failure detection happens through lease expiration and recovery latency is determined by lease length. 

\noindent\textbf{Straggler Mitigation:} The lease mechanism also enables straggler mitigation by default. If a worker stalls or is slow, it can fail to renew a lease before it expires. In this case, a task can be executed by multiple workers. The runtime limit imposed by serverless system act as a global limit for the amount of times a worker can renew their lease, after which the worker will terminate and the task will be handed to a different worker. Because all tasks are idempotent, this has no effect on the correctness, and can speed up execution. {\name} does not require a task queue to have strong guarantees such as exactly-once, in-order delivery, as long as the queue can deliver each task at least once. Such weak ``at-least once delivery'' guarantee is provided by most queue services. %

\subsection{Optimizations}
\label{subsec:optimizations}
We next describe optimizations that improve performance by fully utilizing resources of a worker.

\noindent\textbf{Pipelining:}
    Every {\irname} instruction block has three execution phases: read, compute and write. To improve CPU utilization and I/O efficiency, we allow a worker to fetch multiple tasks and run them in parallel. The number of parallel tasks is called {\em pipeline width}. Ideally, with a single-core worker, we can have at most three tasks running in parallel, each doing read, compute and write respectively. With an appropriately chosen block size, we get best utilization when these three phases take approximately same time. We find pipelining to greatly improve overall utilization, and reduce end-to-end completion time when resources are constrained.\\
\noindent\textbf{Auto Scaling:}
    In contrast to the traditional serverless computing model where each new task is assigned a new container, task scheduling and worker management is decoupled in {\name}. This decoupling allows auto-scaling of computing resources for a better cost-performance trade-off. Historically many auto-scaling policies have been explored \cite{roy2011efficient}.
        In {\name}, we adopt a simple auto-scaling heuristic and find it achieves good utilization while keeping job completion time low.\\
    For scaling up, {\name}'s auto-scaling framework tracks the number of pending tasks and periodically increases the number of running workers to match the pending tasks with a scaling factor $sf$. For instance, let $sf = 0.5$, when there are $100$ pending tasks, $40$ running workers, we launch another $100*0.5-40 = 10$ workers. If pipeline width is not 1, {\name} also factors in pipeline width. For scaling down, {\name} uses an expiration policy where each worker shuts down itself if no task has been found for the last $T_{timeout}$ seconds. At equilibrium, the number of running workers is $sf$ times the number of pending tasks. All of the auto-scaling logic is handled by the ``provisioner" in Figure \ref{fig:execution}.
\begin{table}[!t]
\begin{center}
\setlength{\tabcolsep}{3pt}
\begin{tabular}{lcccccc}
\toprule
Algorithm & \begin{tabular}[c]{@{}c@{}} ScaLAPACK\\(sec) \end{tabular}  &  \begin{tabular}[c]{@{}c@{}} {\name} \\ (sec) \end{tabular} &  \begin{tabular}[c]{@{}c@{}}  Slow \\ down \end{tabular} \\
\midrule
       SVD &                        57,919 &                       77,828 &                 1.33x        \\
       QR &                        3,486 &                       25,108 &                 7.19x  \\
      GEMM &                         2,010 &                        2,670 &                 1.33x  \\
  Cholesky &                         2,417 &                        3,100 &                 1.28x  \\
\bottomrule
\end{tabular}

\caption{\label{tab:algcomp} A comparison of ScaLAPACK vs {\name} execution time across algorithms when run on a square matrix with N=256K}
\end{center}
\end{table}

\section{Evaluation}
\label{sec:exps}
We evaluate {\name} on 4 linear algebra algorithms Matrix Multiply (GEMM), QR Decomposition (QR) , Singular Value Decomposition (SVD) \footnote{Only the reduction to banded form is done in parallel for the SVD} and Cholesky Decomposition (Cholesky).  All of the algorithms have computational complexity of $\mathcal{O}(N^{3})$ but differ in their data access patterns.
For all four algorithms we compare to ScalaPACK, an industrial strength Fortran library designed for high performance, distributed dense linear algebra. We then break down the underlying communication overheads imposed by the serverless computing model. 
We also do a detailed analysis of the scalability and fault tolerance of our system using the Cholesky decomposition. We compare our performance to Dask~\cite{rocklin2015dask}, a python-based fault-tolerant library that supports distributed linear algebra.
Finally, we evaluate optimizations in {\name} and how they affect performance and adaptability.

\begin{figure}[!t]
    \begin{subfigure}[!b]{0.49\columnwidth}
            \includegraphics[width=\textwidth]{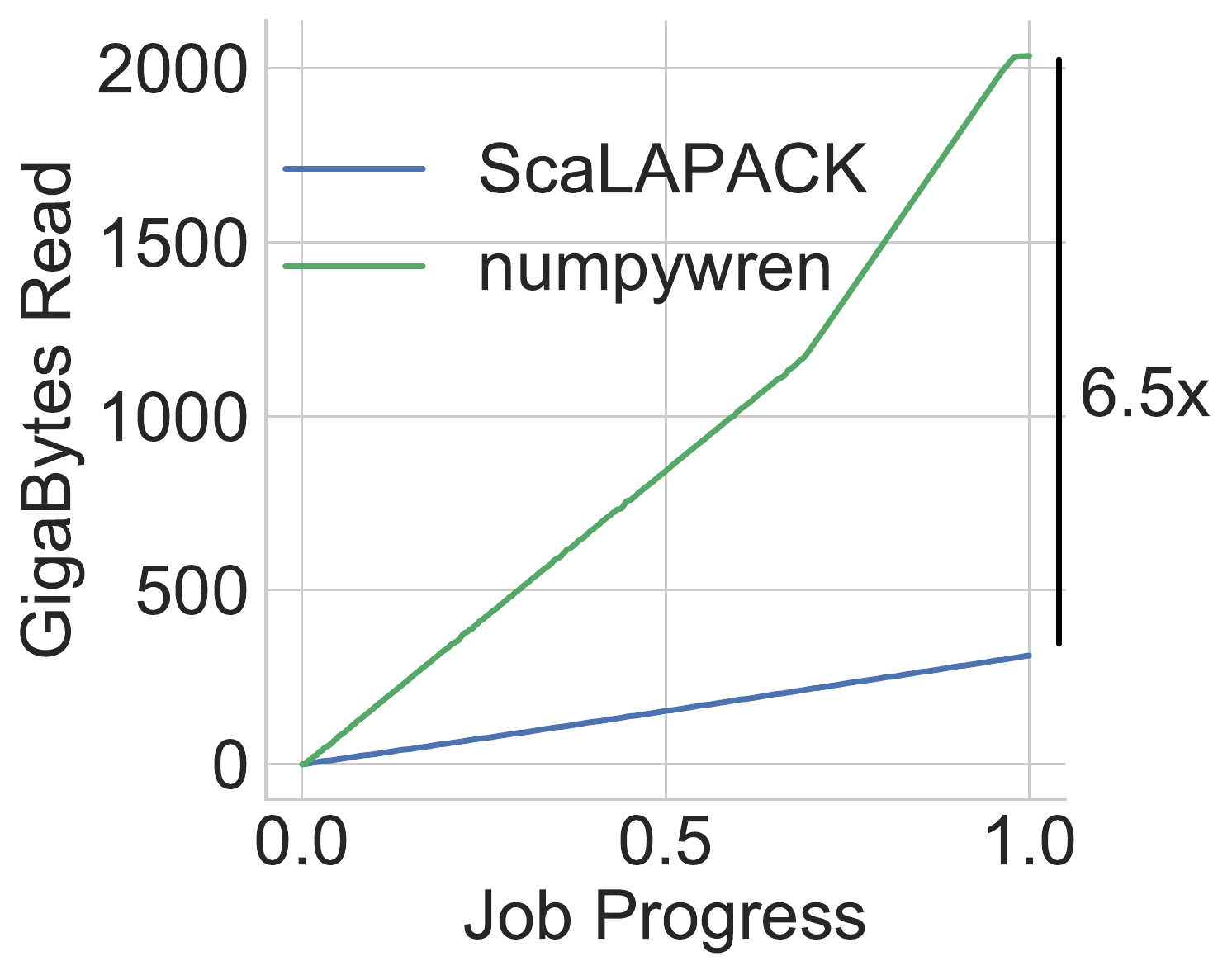}
            \caption{GEMM}
    \end{subfigure}
    \begin{subfigure}[!b]{0.49\columnwidth}
            \includegraphics[width=\textwidth]{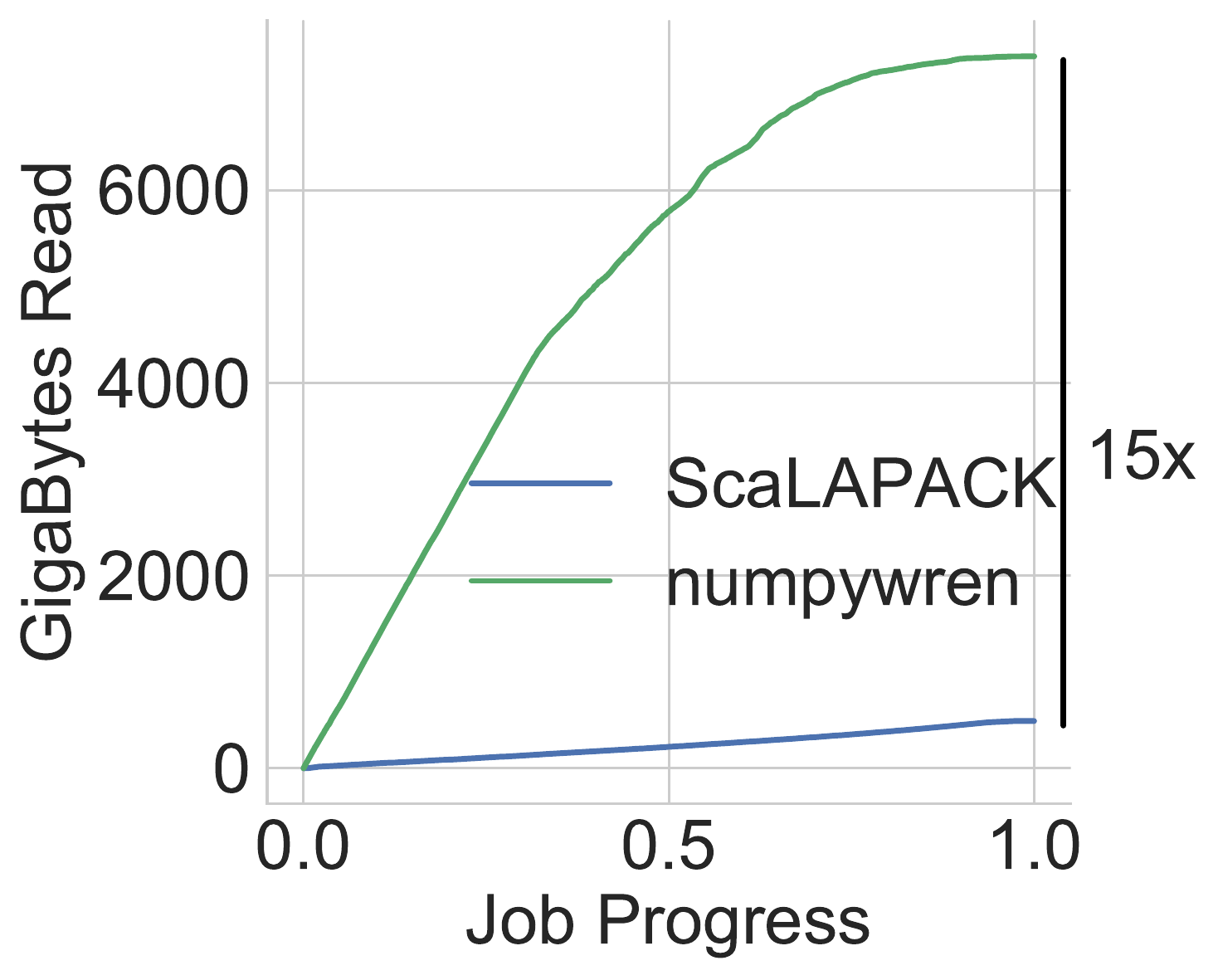}
             \caption{QR}
    \end{subfigure}
    \caption{\label{fig:read_comp} Comparing network bytes for GEMM and QR}
 
\end{figure}

\begin{center}
\begin{table}[!t]
\setlength{\tabcolsep}{5pt}
\begin{tabular}{ccccccc}
\toprule
Algorithm  &  \begin{tabular}[c]{@{}c@{}} {\name} \\ (core-secs) \end{tabular} & \begin{tabular}[c]{@{}c@{}}  ScaLAPACK \\ (core-secs) \end{tabular} &  \begin{tabular}[c]{@{}c@{}} Resource \\ saving \end{tabular} \\
\midrule
       SVD &                          8.6e6 &  2.1e7 &  2.4x  \\
       QR &                          3.8e6 & 1.3e6 & 0.31x  \\
      GEMM &                            1.9e6  &  1.4e6   & 0.74x \\
  Cholesky &                         3.4e5    & 4.3e5 &    1.26x\\
\bottomrule
\end{tabular}
\caption{\label{tab:algcomp_corehours} A comparison of ScaLAPACK vs {\name} total CPU time (in core-secs) across algorithms run on a 256K size square matrix. Resource saving is defined as
$\frac{\text{ScaLAPACK core-secs}}{\text{\name{} core-secs}}$.
}

\end{table}
\end{center}

\vspace{-0.1in}
\subsection{Setup}
\noindent\textbf{Implementation.} Our implementation of {\name} is around 6000 lines of Python code and we build on the Amazon Web Service (AWS) platform. For our runtime state store we use Redis, a key-value store offered by ElasticCache. Though ElasticCache is a provisioned (not ``serverless") service we find that using a single instance suffices for all our workloads. %
We used Amazon's simple queue service (SQS) for the task queue, Lambda for function execution, and S3 for object storage.
We run ScaLAPACK and Dask on \verb|c4.8xlarge|\footnote{60 GB of memory, 18 Physical Cores, 10 GBit network link} instances. 
To obtain a fair comparison with ScaLAPACK and Dask, when comparing to other frameworks we run {\name} on a ``emulated" Lambda environment on the same EC2 instances used for other systems~\footnote{After imposing all the constraints enforced by AWS Lambda in this emulated environment (memory, disk, runtime limits), we found no performance difference between real Lambda and our emulation.}. We chose the number of instances for each problem size by finding the minimum number of instances such that ScaLAPACK could complete the algorithm successfully.

\subsection{System Comparisons}
We first present end-to-end comparison of {\name} to ScaLAPACK on four widely used dense linear algebra methods in Table~\ref{tab:algcomp}.
We compare ScaLAPACK to {\name} when operating on square matrices of size $256K$. 

In table \ref{tab:algcomp} we see that the constraints imposed by the serverless environment lead to a performance penalty between 1.3x to 7x in terms of wall clock time. The difference in the runtime of QR is particularly large, we note that this is primarily due to the high communication penalty our system incurs due to the constraints imposed by the serverless environment. 

In Figure~\ref{fig:read_comp} we compare the number of bytes read over the network by a single machine for two algorithms: GEMM and QR decomposition.
We see that the amount of bytes read by {\name} is always greater than ScaLAPACK. This is a direct consequence of each task being stateless, thus all its arguments must be read from a remote object store.  Moreover we see that for QR decomposition and GEMM, ScaLAPACK reads 15x and 6x less data respectively than {\name}. We discuss future work to address this in Section~\ref{sec:discussion}. 

In Table \ref{tab:algcomp_corehours} we compute the total amount of core-seconds used by {\name} and ScaLAPACK. 
For ScaLAPACK the core-seconds is the total amount of cores multiplied by the wall clock runtime. 
For {\name} we calculate how many cores were actively working on tasks at any given point in time during computation to calculate the total core-seconds. 
For algorithms such as SVD and Cholesky that have variable parallelism, while our wall clock time is comparable (within a factor of 2), we find that {\name} uses 1.26x to 2.5x less resources. However for algorithms that have a fixed amount of parallelism such as GEMM, the excess communication performed by {\name} leads to a higher resource consumption.

\begin{figure*}[!t]
\begin{subfigure}[!b]{0.35\textwidth}
    \centering
    \includegraphics[width=\textwidth]{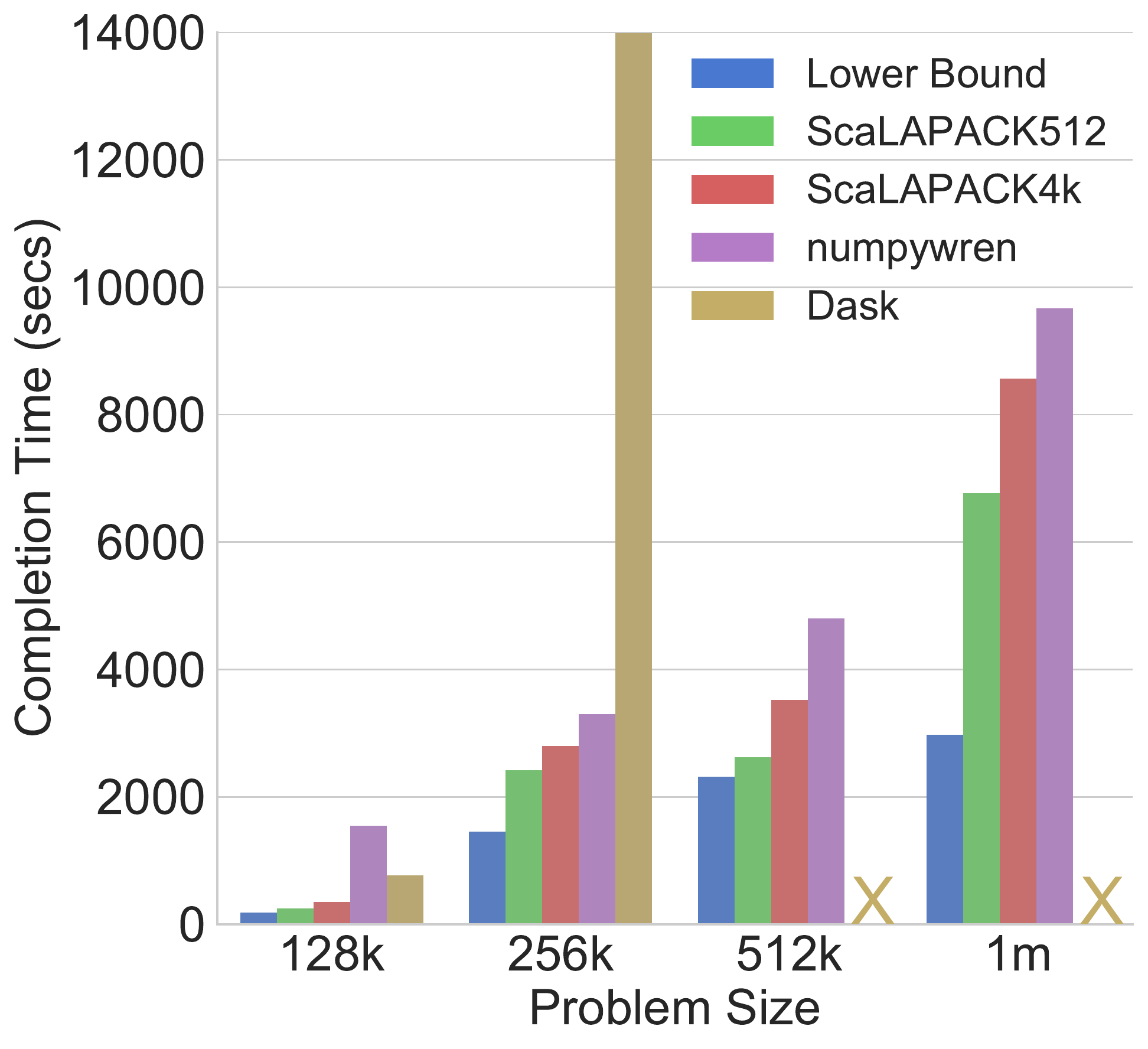}
    \caption{}
    \label{fig:comparison}
\end{subfigure}
\hspace{0.1in}
\begin{subfigure}[!b]{0.3\textwidth}
    \centering
    \includegraphics[width=\textwidth]{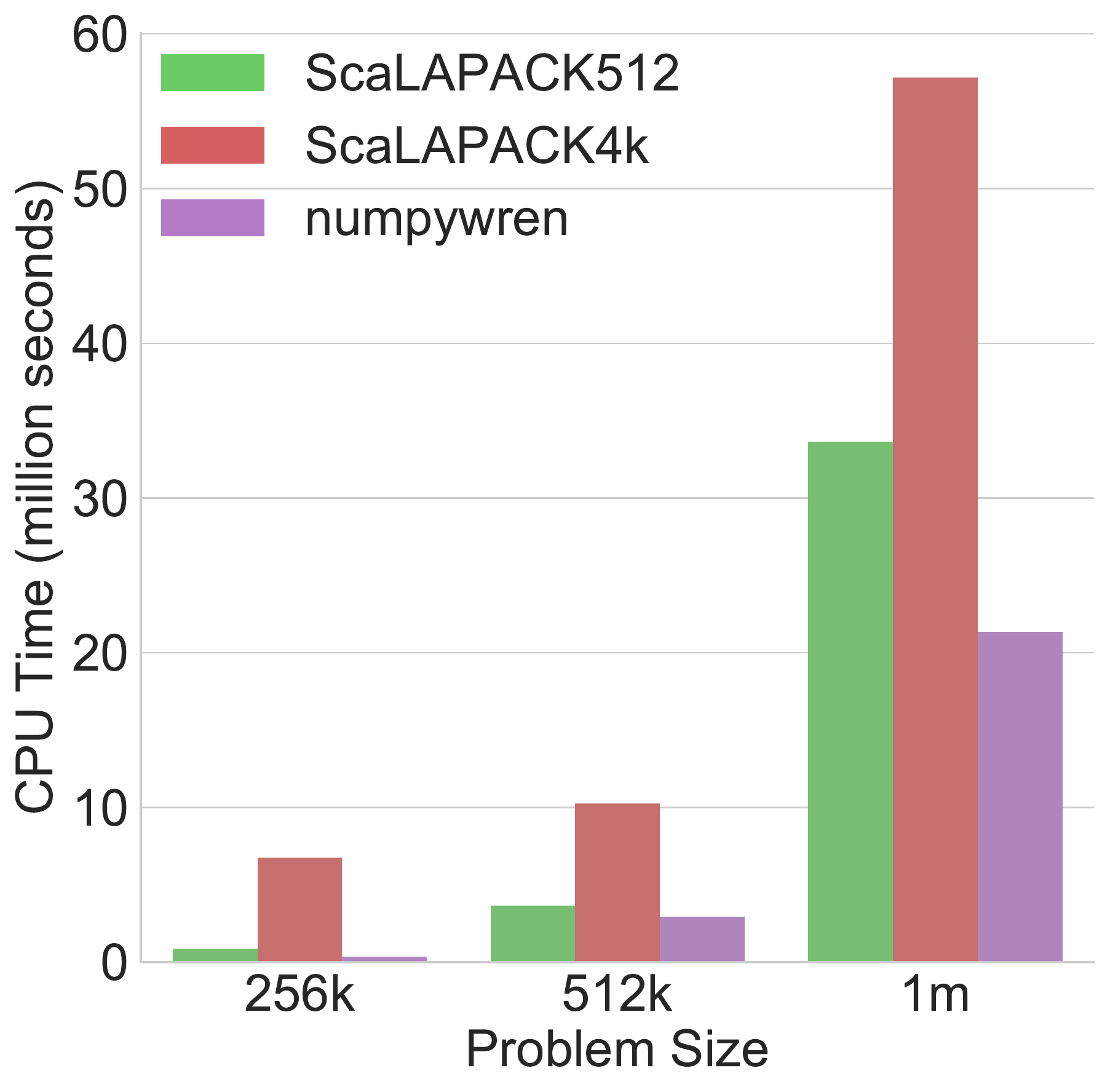}
    \caption{}
    \label{fig:comparison_core_hours}
\end{subfigure}
\begin{subfigure}[!t]{0.3\textwidth}
    \centering
    \includegraphics[width=\columnwidth]{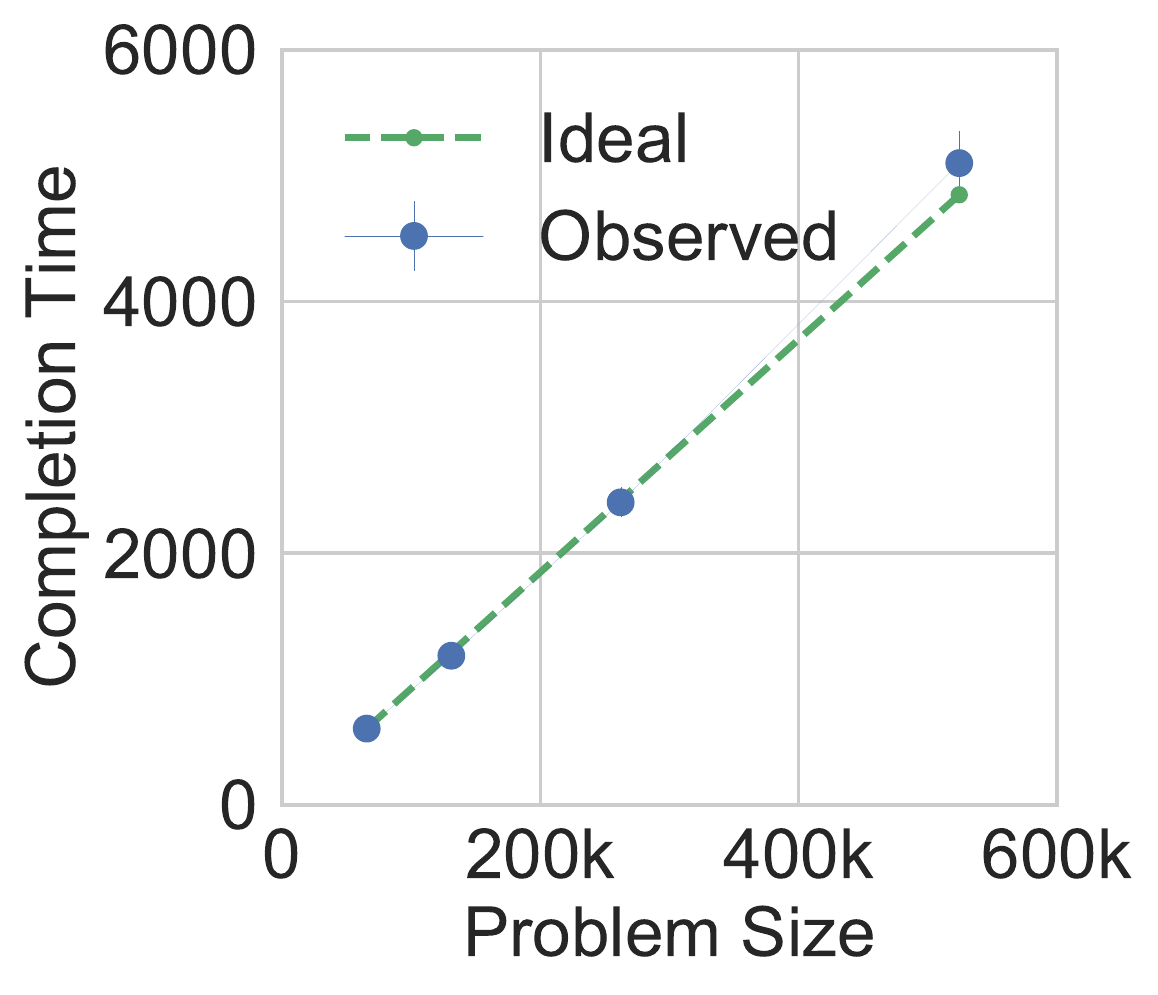}
    \caption{}
    \label{fig:weakscaling}
\end{subfigure}
 \vspace{-0.15in}
\caption{a) Completion time on various problem sizes when {\name} is run on same setup as ScaLAPACK and Dask.   b) Total execution core-seconds for Cholesky when the {\name}, ScaLAPACK, and Dask are optimized for utilization. c) Weak scaling behavior of {\name}. Error bars show minimum and maximum time.}
\end{figure*}

\subsection{Scalability}
We next look at scalability of {\name} and use the Cholesky decomposition study performance and utilization as we scale.
For ScaLAPACK and Dask, we start with 2 instances for the smallest problem size. We scale the number of instances by 4x for a 2x increase in matrix dimension to ensure that the problem fits in cluster memory.
Figure ~\ref{fig:comparison} shows the completion time when running Cholesky decomposition on each framework, as we increase the problem size. Similar to {\name}, ScaLAPACK has a configurable block size that affects the coarseness of local computation.  We report completion time for two different block sizes (4K and 512) for ScaLAPACK in Figure~\ref{fig:comparison}. We use a block size of 4K for {\name}. To get an idea of the communication overheads, we also plot a lower bound on completion time based on the clock-rate of the CPUs used. 

From the figure we see that {\name} is 10 to 15\% slower than ScaLAPACK-4K and 36\% slower than ScaLAPACK-512. Compared to ScaLAPACK-4K, we perform more communication due to the stateless nature of our execution. ScaLAPACK-512 on the other hand has 64x more parallelism but correspondingly the blocks are only 2MB in size and the small block size does not affect the MPI transfers. While {\name} is 50\% slower than Dask at smaller problem sizes, this is because ask execution happens on one machine for small problems avoiding communication. However on large problem sizes, Dask spends a majority of its time serializing and deserializing data and fails to complete execution for the 512k and 1M matrix sizes.

\begin{figure*}[!t]
\centering
\begin{subfigure}[!t]{0.45\textwidth}
    \centering
    \includegraphics[width=0.75\columnwidth]{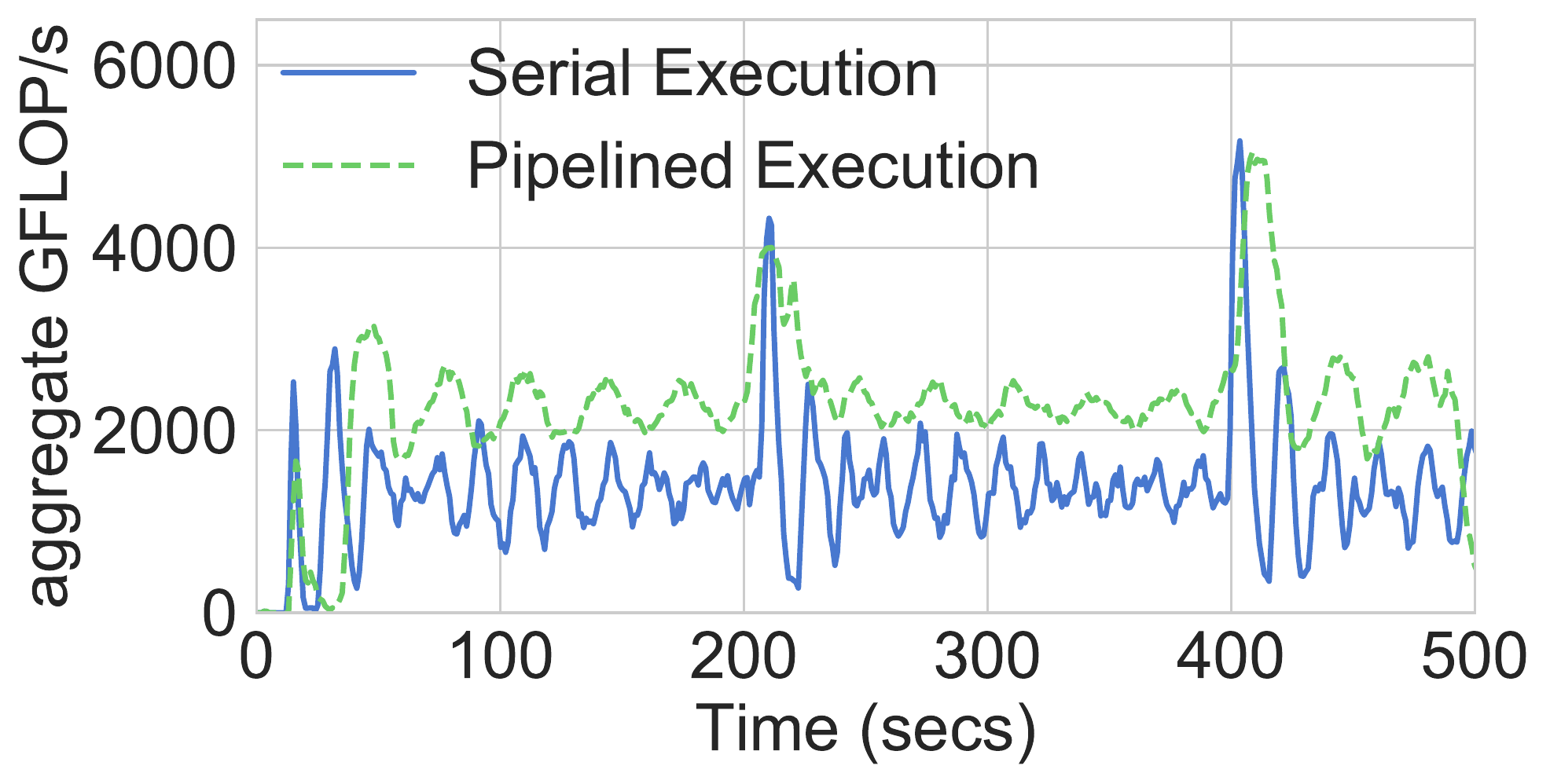}
    \vspace{-0.12in}
    \caption{}
    \label{fig:pipeline}
\end{subfigure}
\hspace{0.1in}
\begin{subfigure}{0.45\textwidth}
    \centering
    \includegraphics[width=0.75\columnwidth]{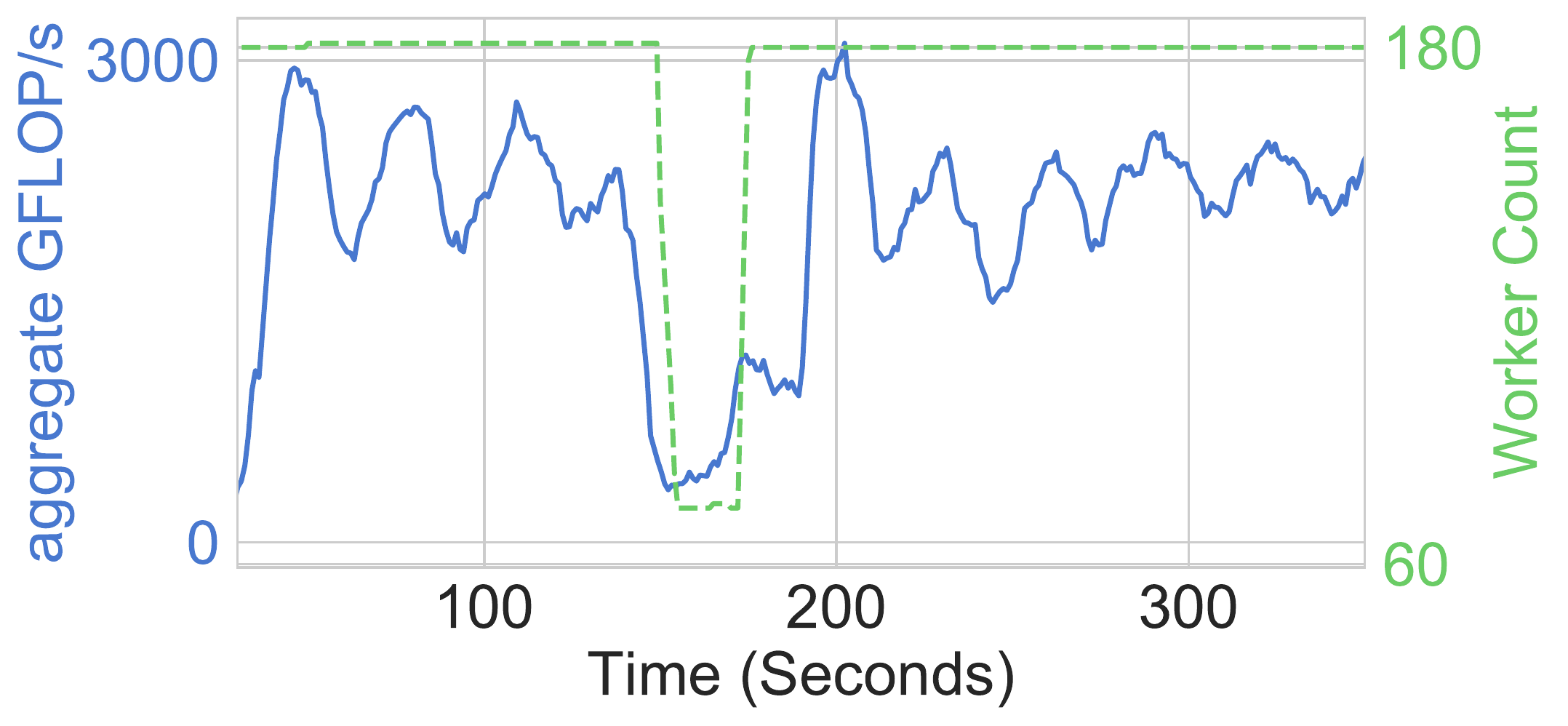}
    \vspace{-0.12in}
    \caption{}
    \label{fig:failures_workers}
\end{subfigure}
\vspace{-0.1in}
\caption{a) Runtime profile with and without pipelining. b) Graceful degredation and recovery of system performance with failure of 80\% of workers.}
 \vspace{-0.1in}
\end{figure*}

\noindent\textbf{Weak Scaling.} In Figure~\ref{fig:weakscaling} we focus on the weak-scaling behavior of {\name}.  Cholesky decomposition has an algorithmic complexity of $O(N^{3})$ and a maximum parallelism of $O(N^{2})$, so we increase our core count quadratically from 57 to 1800 as we scale the problem from 65k to 512k. We expect our ideal curve (shown by the green line in Figure~\ref{fig:weakscaling}) to be a diagonal line. We see that our performance tracks the ideal behavior quite well despite the extra communication overheads incurred.

\noindent\textbf{Utilization.} We next look at how resource utilization varies with scale.
We compare aggregate core-hours in Figure ~\ref{fig:comparison_core_hours} for different problem sizes. In this experiment we configured all three frameworks, ScaLAPACK, Dask and {\name} to minimize total resources consumed. We note that for ScaLAPACK and Dask this is often the minimum number of machines needed to fit the problem in memory.
Compared to ScaLAPACK-512 we find that {\name} uses $20$\% to $33$\%  lower core hours. Disaggregated storage allows {\name} to have the flexibility to run with 4x \textbf{less} cores but increases completion time by 3x. In contrast to {\name}, cluster computation frameworks need a minimum resource allocation to fit the problem in memory, thus such a performance/resource consumption trade-off is not possible on Dask or ScaLAPACK.

\begin{figure*}
\begin{subfigure}[!t]{0.33\textwidth}
    \centering
    \includegraphics[width=\columnwidth]{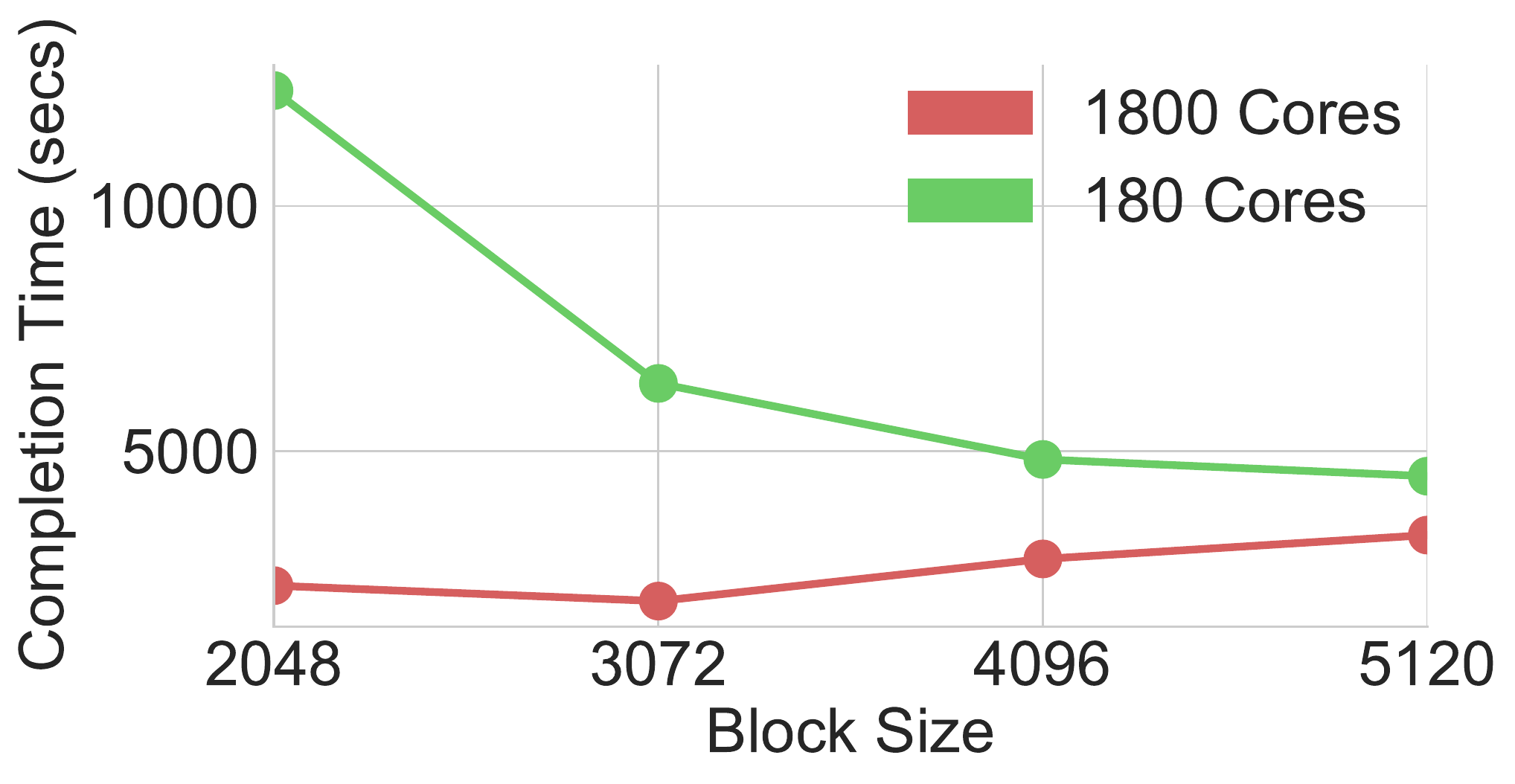}
    \caption{\label{fig:blocksize_exp}}
        \vspace{-0.15in}
\end{subfigure}
\begin{subfigure}[!t]{0.33\textwidth}
    \centering
    \includegraphics[width=\columnwidth]{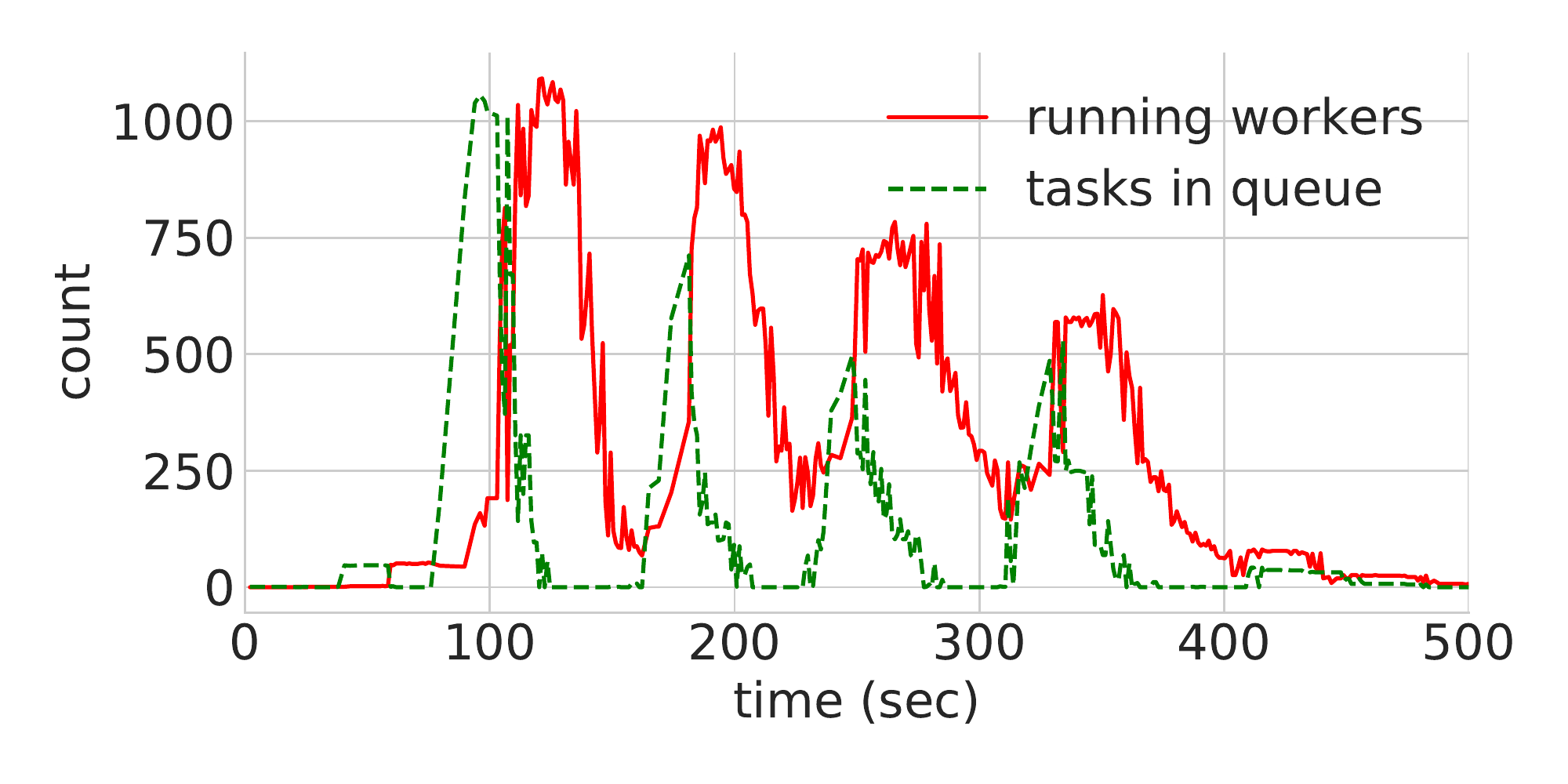}
    \caption{}
    \label{fig:autoscaling_queue}
        \vspace{-0.15in}
\end{subfigure}
\begin{subfigure}[!t]{0.33\textwidth}
    \centering
    \includegraphics[width=\columnwidth]{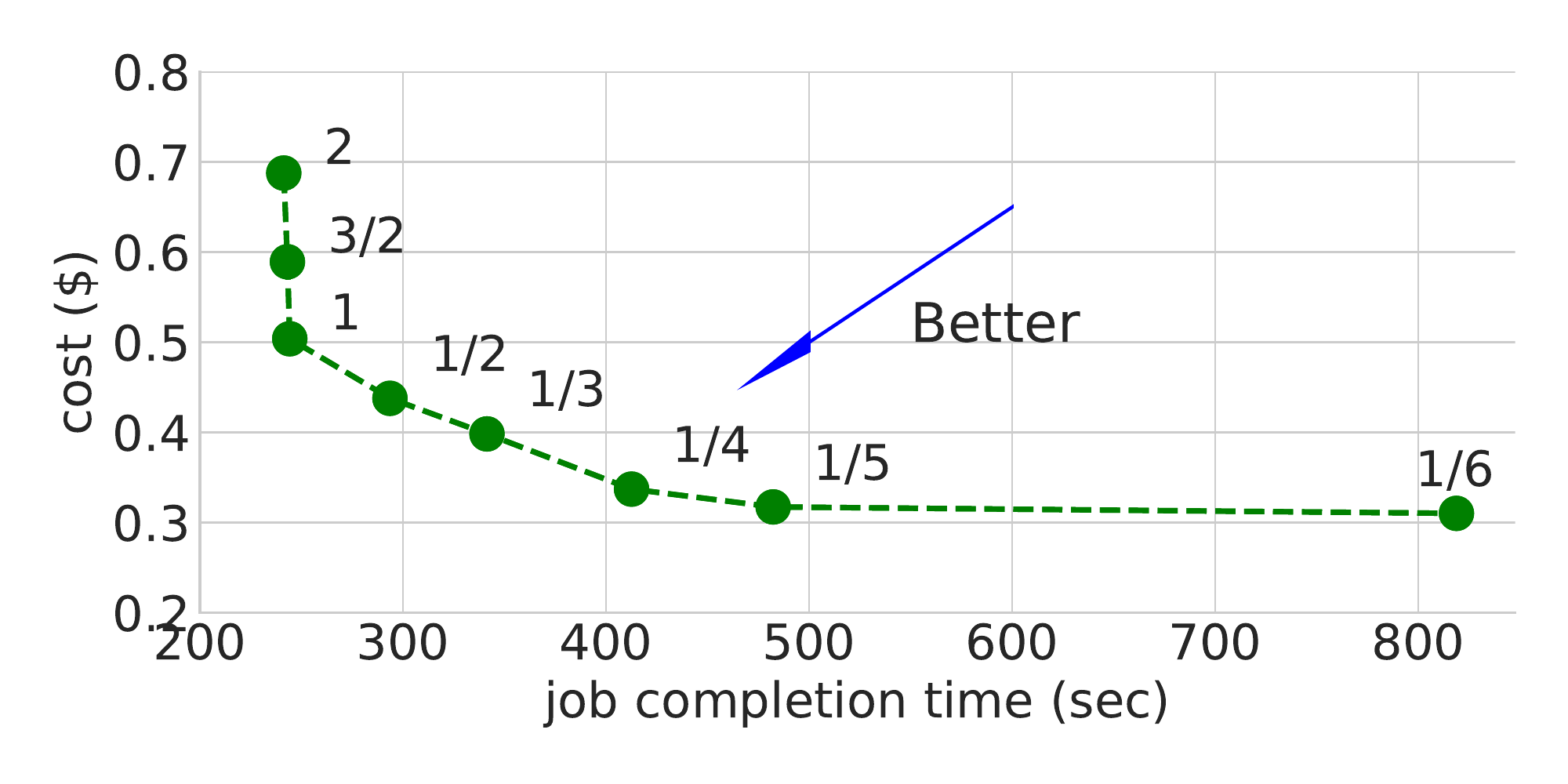}
        \vspace{-0.25in}
    \caption{}
    \label{fig:autoscaling_cost_trade_off}
    \vspace{-0.1in}
\end{subfigure}
\caption{a) Effect of block size on completion time b) Our auto-scaling policy in action. The number of workers increases as the task queue builds up, decreases as the queue is being cleared c) Cost performance trade-off when varying auto-scaling factor (as labeled next to the data points)}
\vspace{-0.1in}
\end{figure*}

\begin{table}
\setlength{\tabcolsep}{1.25pt}
\begin{tabular}{cccccccc}
\toprule
      N &   \begin{tabular}[c]{@{}c@{}} Full \\  DAG \\  Time (s) \end{tabular}  &  \begin{tabular}[c]{@{}c@{}} LambdaPack \\ time (s) \end{tabular} & \begin{tabular}[c]{@{}c@{}} DAG  \\ Size \\ (\# nodes) \end{tabular} &   \begin{tabular}[c]{@{}c@{}} Expanded\\ DAG \\  (MB) \end{tabular} &    \begin{tabular}[c]{@{}c@{}} Compiled \\ Program \\ (MB)  \end{tabular}  \\
\midrule
    65k &                       3.56&                  0.019 &              4k &                     0.6 &                    0.027 \\
   128k &                        4.02 &                  0.027 &            32k &                    4.6 &                    0.027  \\
  256k &                       12.57 &                  0.065 &            256k &                   36.3 &                   0.027 \\
  512k &                       49.0 &                  0.15 &           2M &                  286  &                    0.027 \\
 1M &                      450 &                  0.44 &          16M &                 2270  &                    0.027 \\
\bottomrule
\end{tabular}
\centering
\caption{\label{tab:compression} Benefits of {\irname} analysis in terms of program size and time to enumerate DAG dependencies.}
\end{table}

\subsection{Optimizations and Adaptability}
We next evaluate optimizations in {\name} (Section~\ref{sec:arch}) and how those affect performance and adapatbility.\\
\noindent\textbf{Pipelining Benefits.}
We measured the benefits of pipelining using Cholesky decomposition of a matrix of size $256K$. 
We see that pipelining drastically improves the resource utilization profile as shown in Figure ~\ref{fig:pipeline}. The \textit{average flop rate} on a 180 core cluster is 40\% higher with pipelining enabled.\\
\noindent\textbf{Fault Recovery.}
We next measure performance of {\name} under intermittent failures of the cloud functions. Failures can be common in this setting as cloud functions can get preempted or slow down due to contention. 
In the experiment in Figure~\ref{fig:failures_workers} we start with 180 workers and after 150 seconds, we inject failures in 80\% of the workers.
The disaggregation of resources and the fine grained computation performed by our execution engine leads to a performance penalty
linear in the amount of workers that fail. Using the autoscaling technique discussed in~\ref{subsec:optimizations}, Figure ~\ref{fig:failures_workers} also shows that we can replenish the worker pool to the original size in 20 seconds. We find there is an extra 20 second delay before the computation picks back up due the the startup communication cost of reading program arguments from the object store.

\noindent\textbf{Auto-scaling.}
Figure~\ref{fig:autoscaling_queue} shows our auto-scaling policy in action. We ran the first 5000 instructions of a 256k Cholesky solve on AWS Lambda with $sf = 1.0$ (as mentioned in subsection ~\ref{subsec:optimizations}) and pipeline width = 1. We see that {\name} adapts to the dynamic parallelism of the workload.
Another important question is how to set the parameters, i.e., scaling factor $sf$ and $T_{timeout}$. We use simple heuristics and empirical experiments to decide these two parameters and leave more rigorous investigation for future work. We set $T_{timeout} = 10s$, which is the average start-up latency of a worker. For $sf$, we want to make sure that when a new worker (started during scaling up) becomes ready, the task queue should not be completely empty, so the worker can be utilized. Figure~\ref{fig:autoscaling_cost_trade_off} shows the trade-off between cost-vs-completion time as we vary $sf$. From the figure we see that as $sf$ decreases we waste fewer resources but the completion time becomes worse. At higher values of $sf$ the job finishes faster but costs more. Finally we see that there are a range of values of $sf$ (1/4, 1/3, 1/2, 1) that balance the cost and execution time. Outside of the range, either there are always tasks in the queue, or overly-aggressive scaling spawns workers that do not execute any tasks. As described in Section~\ref{subsec:optimizations}, the balanced range is determined by worker start-up latency, task graph, execution time and pipeline width.%
\\
\noindent\textbf{DAG Compression.}
In Table \ref{tab:compression} we measure the {\irname}'s ability to express large program graphs with constant space, and moreover that we can compile such programs quickly. This is crucial for efficient execution since memory is scarce in the serverless environment, and distributing a large program DAG to each worker can dramatically affect performance. We see that as matrix sizes grow to 1Mx1M the DAG takes over 2 GB of memory {\irname} lowers this to 2 KB making it feasible to execute on large matrices.\\
\noindent\textbf{Blocksize}
\label{subsec:blocksize_disc}
A parameter that is of interest in performance tuning of distributed linear algebra algorithms is the \textit{block size} which defines the coarseness of computation. We evaluate the effect of block size on completion time in Figure~\ref{fig:blocksize_exp}. We run the same workload (a $256K$ Cholesky decomposition) at two levels of parallelism, 180 cores and 1800 cores. We see that in the 180 core case, larger block size leads to significantly faster completion time as each task performs more computation and can hide communication overheads. With higher parallelism, we see that the largest block size is slowest as it has the least opportunity to exploit the parallelism available. However, we also see that the smallest block size (2048) is affected by latency overheads in both regimes.

\section{Related Work}

\noindent\textbf{Distributed Linear Algebra Libraries} Building distributed systems for linear algebra has long been an active area of research. Initially, this was studied in the context of High Performance Computing (HPC), where frameworks like ScaLAPACK~\cite{scalapack}, DPLASMA~\cite{dplasma} and Elemental~\cite{elemental} run on a multi-core, shared-memory architecture with high performance network interconnect. However, on-demand access to a HPC cluster can be difficult. While one can run ScaLAPACK or DPLASMA in the cloud, it is undesirable due to their lack of fault tolerance.
On the other hand, with the wide adoption of MapReduce or BSP-style data analytics in the cloud, a number of systems have implemented linear algebra libraries~\cite{systemml, mllib, mahout, marlin, hama}. However, the BSP-style programming API is ill-suited for expressing the fine-grained dependencies in linear algebra algorithms, and imposing global synchronous barriers often greatly slows down a job. Thus not surprisingly, none of these systems~\cite{systemml, mllib, mahout, marlin} have an efficient implementation of distributed Cholesky decomposition that can compare with {\name} or ScaLAPACK. The only dataflow-based system that supports fine grained dependencies is MadLINQ~\cite{madlinq}. {\name} differs from MadLINQ in that it is designed for a serverless architecture and achieves recomputation-free failure (since the previously computed blocks will remain in the object store) recovery by leveraging resource disaggregation, compared to MadLINQ where lost blocks need to be recomputed during recovery. SystemML \cite{systemml} takes a similar approach to \irname{} in providing a high level framework for  numerical computation, however they target a BSP backend and focus on machine learning algorithms as opposed to linear algebra primitives.

\noindent\textbf{Serverless Frameworks}: The paradigm shift to serverless computing has brought new innovations to many traditional applications. One predominant example is SQL processing, which is now offered in a serverless mode by many cloud providers~\cite{bigquery, athena, glue, redshiftspectrum}. Serverless general computing platforms (OpenLambda~\cite{openlambda}, AWS Lambda, Google Cloud Functions, Azure Functions, etc.) have led to new computing frameworks~\cite{aws-lambda-mapred, fouladi2017encoding, jonas2017occupy}. Even a complex analytics system such as Apache Spark has been ported to run on AWS Lambda~\cite{sparkonlambda}. However, none of the previous frameworks deal with complex communication patterns across stateless workers. {\name} is, to our knowledge, the first large-scale linear algebra library that runs on a serverless architecture.

\noindent\textbf{Auto-scaling and Fault Tolerance} Efforts that add fault tolerance to ScaLAPACK has so far demonstrated to incur significant performance overhead~\cite{bland2012checkpoint}. For almost all BSP and dataflow systems\cite{ciel, dryad, ray}, recomputation is required to restore stateful workers or datasets that have not been checkpointed. MadLINQ~\cite{madlinq} also uses dependency tracking to minimize recomputation for its pipelined execution. In contrast, {\name} uses a serverless computing model where fault tolerance only requires re-executing failed tasks and no recomputation is required. {\name}'s failure detection is also different  and we use a lease-based mechanism. The problem of auto-scaling cluster size to fit dynamic workload demand has been both studied~\cite{autoscalingtheory} and deployed by many cloud vendors. However, due to the relatively high start-up latency of virtual machines, its cost-saving capacity has been limited. {\name} exploits the elasticity of serverless computing to achieve better cost-performance trade-off.%

\section{Discussion and Future Work}
\label{sec:discussion}
\noindent\textbf{Collective Communication and Colocation.} One of the main drawbacks of the serverless model is the high communication needed due to the lack of locality and efficient broadcast primitives. One way to alleviate this would be to have coarser serverless executions (e.g., 8 cores instead of 1) that process larger portions of the input data.  Colocation of lambdas could also achieve similar effects if the colocated lambdas could efficiently share data with each other. Finally, developing services that provide efficient collective communication primitives like broadcast will also help address this problem.

\noindent\textbf{Higher-level libraries.} The high level interface in {\name} paves way for easy algorithm design and we believe modern convex optimization solvers such as CVXOPT can use {\name} to scale to much larger problems. Akin to Numba \cite{numba} we are also working on automatically translating \verb|numpy| code directly into {\irname} instructions than can be executed in parallel. 

In conclusion, we have presented {\name}, a distributed system for executing large-scale dense linear algebra programs via stateless function executions.
We show that the serverless computing model can be used for computationally intensive programs with complex communication routines while providing ease-of-use and seamless fault tolerance, through analysis of the intermediate \irname{} language. Furthermore, the elasticity provided by serverless computing allows our system to dynamically adapt to the inherent parallelism of common linear algebra algorithms. As datacenters continue their push towards disaggregation, platforms like {\name} open up a fruitful area of research for applications that have long been dominated by traditional HPC. 
\section{Acknowledgements}
This research is supported in part by ONR awards N00014-17-1-2191, N00014-17-1-2401, and N00014-18-1-2833, the DARPA Assured Autonomy (FA8750-18-C-0101) and Lagrange (W911NF-16-1-0552) programs, Amazon AWS AI Research Award, NSF CISE Expeditions Award CCF-1730628 and gifts from
Alibaba, Amazon Web Services, Ant Financial, Arm, CapitalOne, Ericsson, Facebook, Google, Huawei, Intel,
Microsoft, Scotiabank, Splunk and VMware as well as by NSF grant DGE-1106400. \\

We would like to thank Horia Mania, Alyssa Morrow and Esther Rolf for helpful comments while writing this paper.

{
  \footnotesize
  \bibliographystyle{acm}
  \bibliography{cites}

\begin{thebibliography}{10}

\bibitem{plasma}
{\sc Agullo, E., Demmel, J., Dongarra, J., Hadri, B., Kurzak, J., Langou, J.,
  Ltaief, H., Luszczek, P., and Tomov, S.}
\newblock Numerical linear algebra on emerging architectures: The plasma and
  magma projects.
\newblock In {\em Journal of Physics: Conference Series\/} (2009), vol.~180,
  IOP Publishing, p.~012037.

\bibitem{anderson2011communication}
{\sc Anderson, M., Ballard, G., Demmel, J., and Keutzer, K.}
\newblock Communication-avoiding qr decomposition for gpus.
\newblock In {\em Parallel \& Distributed Processing Symposium (IPDPS), 2011
  IEEE International\/} (2011), IEEE, pp.~48--58.

\bibitem{athena}
{Amazon Athena}.
\newblock \url{http://aws.amazon.com/athena/}.

\bibitem{aws-lambda-mapred}
{Serverless Reference Architecture: MapReduce}.
\newblock \url{https://github.com/awslabs/lambda-refarch-mapreduce}.

\bibitem{awshpc}
{Amazon AWS High Performance Clusters}.
\newblock \url{https://aws.amazon.com/hpc}.

\bibitem{azurehpc}
{Microsoft Azure High Performance Computing}.
\newblock
  \url{https://azure.microsoft.com/en-us/solutions/high-performance-computing}.

\bibitem{ballard2010communication}
{\sc Ballard, G., Demmel, J., Holtz, O., and Schwartz, O.}
\newblock Communication-optimal parallel and sequential cholesky decomposition.
\newblock {\em SIAM Journal on Scientific Computing 32}, 6 (2010), 3495--3523.

\bibitem{ballard2011minimizing}
{\sc Ballard, G., Demmel, J., Holtz, O., and Schwartz, O.}
\newblock Minimizing communication in numerical linear algebra.
\newblock {\em SIAM Journal on Matrix Analysis and Applications 32}, 3 (2011),
  866--901.

\bibitem{bigquery}
{Google BigQuery}.
\newblock \url{https://cloud.google.com/bigquery/}.

\bibitem{scalapack}
{\sc Blackford, L.~S., Choi, J., Cleary, A., Petitet, A., Whaley, R.~C.,
  Demmel, J., Dhillon, I., Stanley, K., Dongarra, J., Hammarling, S., Henry,
  G., and Walker, D.}
\newblock Scalapack: A portable linear algebra library for distributed memory
  computers - design issues and performance.
\newblock In {\em Proceedings of ACM/IEEE Conference on Supercomputing\/}
  (1996).

\bibitem{bland2012checkpoint}
{\sc Bland, W., Du, P., Bouteiller, A., Herault, T., Bosilca, G., and Dongarra,
  J.}
\newblock A checkpoint-on-failure protocol for algorithm-based recovery in
  standard mpi.
\newblock In {\em European Conference on Parallel Processing\/} (2012),
  Springer, pp.~477--488.

\bibitem{systemml}
{\sc Boehm, M., Dusenberry, M.~W., Eriksson, D., Evfimievski, A.~V., Manshadi,
  F.~M., Pansare, N., Reinwald, B., Reiss, F.~R., Sen, P., Surve, A.~C.,
  et~al.}
\newblock Systemml: Declarative machine learning on spark.
\newblock {\em Proceedings of the VLDB Endowment 9}, 13 (2016), 1425--1436.

\bibitem{dplasma}
{\sc Bosilca, G., Bouteiller, A., Danalis, A., Faverge, M., Haidar, A.,
  Herault, T., Kurzak, J., Langou, J., Lemarinier, P., Ltaief, H., et~al.}
\newblock Flexible development of dense linear algebra algorithms on massively
  parallel architectures with dplasma.
\newblock In {\em Parallel and Distributed Processing Workshops and Phd Forum
  (IPDPSW), 2011 IEEE International Symposium on\/} (2011), IEEE,
  pp.~1432--1441.

\bibitem{dague}
{\sc Bosilca, G., Bouteiller, A., Danalis, A., Herault, T., Lemarinier, P., and
  Dongarra, J.}
\newblock Dague: A generic distributed dag engine for high performance
  computing.
\newblock {\em Parallel Computing 38}, 1-2 (2012), 37--51.

\bibitem{fouladi2017encoding}
{\sc Fouladi, S., Wahby, R.~S., Shacklett, B., Balasubramaniam, K., Zeng, W.,
  Bhalerao, R., Sivaraman, A., Porter, G., and Winstein, K.}
\newblock Encoding, fast and slow: Low-latency video processing using thousands
  of tiny threads.
\newblock In {\em NSDI\/} (2017), pp.~363--376.

\bibitem{gao2016network}
{\sc Gao, P.~X., Narayan, A., Karandikar, S., Carreira, J., Han, S., Agarwal,
  R., Ratnasamy, S., and Shenker, S.}
\newblock Network requirements for resource disaggregation.
\newblock In {\em OSDI\/} (2016), vol.~16, pp.~249--264.

\bibitem{georganas2012communication}
{\sc Georganas, E., Gonzalez-Dominguez, J., Solomonik, E., Zheng, Y., Tourino,
  J., and Yelick, K.}
\newblock Communication avoiding and overlapping for numerical linear algebra.
\newblock In {\em Proceedings of the International Conference on High
  Performance Computing, Networking, Storage and Analysis\/} (2012), IEEE
  Computer Society Press, p.~100.

\bibitem{glue}
{Amazon Glue}.
\newblock \url{https://aws.amazon.com/glue/}.

\bibitem{googlehpc}
{Google Cloud High Performance Computing}.
\newblock
  \url{https://cloud.google.com/solutions/architecture/highperformancecomputing}.

\bibitem{cheriton-leases}
{\sc Gray, C., and Cheriton, D.}
\newblock Leases: An efficient fault-tolerant mechanism for distributed file
  cache consistency.
\newblock In {\em Proceedings of the Twelfth ACM Symposium on Operating Systems
  Principles\/} (1989), SOSP '89, pp.~202--210.

\bibitem{marlin}
{\sc Gu, R., Tang, Y., Tian, C., Zhou, H., Li, G., Zheng, X., and Huang, Y.}
\newblock Improving execution concurrency of large-scale matrix multiplication
  on distributed data-parallel platforms.
\newblock In {\em IEEE Transactions on Parallel \& Distributed Systems\/}
  (2017).

\bibitem{heide2016proximal}
{\sc Heide, F., Diamond, S., Niessner, M., Ragan-Kelley, J., Heidrich, W., and
  Wetzstein, G.}
\newblock Proximal: Efficient image optimization using proximal algorithms.
\newblock {\em ACM Transactions on Graphics (TOG) 35}, 4 (2016), 84.

\bibitem{openlambda}
{\sc Hendrickson, S., Sturdevant, S., Harter, T., Venkataramani, V.,
  Arpaci-Dusseau, A.~C., and Arpaci-Dusseau, R.~H.}
\newblock Serverless computation with openlambda.
\newblock In {\em Proceedings of the 8th USENIX Conference on Hot Topics in
  Cloud Computing\/} (2016), HotCloud'16.

\bibitem{dryad}
{\sc Isard, M., Budiu, M., Yu, Y., Birrell, A., and Fetterly, D.}
\newblock Dryad: distributed data-parallel programs from sequential building
  blocks.
\newblock {\em ACM SIGOPS operating systems review 41}, 3 (2007), 59--72.

\bibitem{jonas2017occupy}
{\sc Jonas, E., Pu, Q., Venkataraman, S., Stoica, I., and Recht, B.}
\newblock Occupy the cloud: distributed computing for the 99\%.
\newblock In {\em Proceedings of the 2017 Symposium on Cloud Computing\/}
  (2017), ACM, pp.~445--451.

\bibitem{mahout}
{Apache Mahout}.
\newblock \url{https://mahout.apache.org}.

\bibitem{autoscalingtheory}
{\sc Mao, M., and Humphrey, M.}
\newblock Auto-scaling to minimize cost and meet application deadlines in cloud
  workflows.
\newblock In {\em Proceedings of 2011 International Conference for High
  Performance Computing, Networking, Storage and Analysis\/} (2011).

\bibitem{mllib}
{\sc Meng, X., Bradley, J., Yavuz, B., Sparks, E., Venkataraman, S., Liu, D.,
  Freeman, J., Tsai, D., Amde, M., Owen, S., Xin, D., Xin, R., Franklin, M.~J.,
  Zadeh, R., Zaharia, M., and Talwalkar, A.}
\newblock Mllib: Machine learning in apache spark.
\newblock {\em Journal of Machine Learning Research 17}, 34 (2016), 1--7.

\bibitem{ray}
{\sc Moritz, P., Nishihara, R., Wang, S., Tumanov, A., Liaw, R., Liang, E.,
  Paul, W., Jordan, M.~I., and Stoica, I.}
\newblock Ray: A distributed framework for emerging ai applications.
\newblock {\em arXiv preprint arXiv:1712.05889\/} (2017).

\bibitem{ciel}
{\sc Murray, D.~G., Schwarzkopf, M., Smowton, C., Smith, S., Madhavapeddy, A.,
  and Hand, S.}
\newblock Ciel: a universal execution engine for distributed data-flow
  computing.
\newblock In {\em Proc. 8th ACM/USENIX Symposium on Networked Systems Design
  and Implementation\/} (2011), pp.~113--126.

\bibitem{numba}
Numba.
\newblock \url{https://numba.pydata.org/}.

\bibitem{parikh2014proximal}
{\sc Parikh, N., Boyd, S., et~al.}
\newblock Proximal algorithms.
\newblock {\em Foundations and Trends in Optimization 1}, 3 (2014), 127--239.

\bibitem{elemental}
{\sc Poulson, J., Marker, B., Van~de Geijn, R.~A., Hammond, J.~R., and Romero,
  N.~A.}
\newblock Elemental: A new framework for distributed memory dense matrix
  computations.
\newblock {\em ACM Transactions on Mathematical Software (TOMS) 39}, 2 (2013),
  13.

\bibitem{madlinq}
{\sc Qian, Z., Chen, X., Kang, N., Chen, M., Yu, Y., Moscibroda, T., and Zhang,
  Z.}
\newblock Madlinq: large-scale distributed matrix computation for the cloud.
\newblock In {\em Proceedings of the 7th ACM European Conference on Computer
  Systems\/} (2012), ACM, pp.~197--210.

\bibitem{redshiftspectrum}
{Amazon Redshift Spectrum}.
\newblock \url{https://aws.amazon.com/redshift/spectrum/}.

\bibitem{rocklin2015dask}
{\sc Rocklin, M.}
\newblock Dask: Parallel computation with blocked algorithms and task
  scheduling.
\newblock In {\em Proceedings of the 14th Python in Science Conference\/}
  (2015).

\bibitem{roy2011efficient}
{\sc Roy, N., Dubey, A., and Gokhale, A.}
\newblock Efficient autoscaling in the cloud using predictive models for
  workload forecasting.
\newblock In {\em Cloud Computing (CLOUD), 2011 IEEE International Conference
  on\/} (2011), IEEE, pp.~500--507.

\bibitem{hama}
{\sc Seo, S., Yoon, E.~J., Kim, J., Jin, S., Kim, J.-S., and Maeng, S.}
\newblock Hama: An efficient matrix computation with the mapreduce framework.
\newblock In {\em CLOUDCOM\/} (2010).

\bibitem{sparkonlambda}
{Apache Spark on AWS Lambda}.
\newblock \url{https://www.qubole.com/blog/spark-on-aws-lambda/}.

\bibitem{tu2017non}
{\sc Tu, S., Boczar, R., Packard, A., and Recht, B.}
\newblock Non-asymptotic analysis of robust control from coarse-grained
  identification.
\newblock {\em arXiv preprint arXiv:1707.04791\/} (2017).

\bibitem{ernest}
{\sc Venkataraman, S., Yang, Z., Franklin, M., Recht, B., and Stoica, I.}
\newblock Ernest: Efficient performance prediction for large-scale advanced
  analytics.
\newblock In {\em NSDI\/} (2016).

\end{thebibliography}
}

\end{document}